\documentclass[a4paper,10pt]{article}

\RequirePackage[OT1]{fontenc}
\RequirePackage{amsthm,amsmath}
\RequirePackage[numbers]{natbib}

\usepackage[normalem]{ulem}
\RequirePackage{lmodern}
\RequirePackage{graphicx}
\graphicspath{{fig/}}
\RequirePackage{amsfonts}
\RequirePackage{amssymb}
\RequirePackage{dsfont} 
\RequirePackage{textcomp}
\RequirePackage[official]{eurosym}
\RequirePackage{fancybox}
\RequirePackage{color}
\definecolor{bleu}{rgb}{0.8,0.9,1}
\definecolor{gris}{gray}{0.8}
\RequirePackage{ulem}
\RequirePackage{enumerate}
\RequirePackage{fancybox}

\newcommand{\dfc}[3]{#1 : #2 \rightarrow #3}

\newcommand{\trans}[1]{{#1}^T}

\newcommand{\1}{\mathds{1}}
\newcommand{\vr}[1]{\left(\begin{array}{c}#1\end{array}\right)}
\newcommand{\Esp}[1]{\mathbb{E}\left[{#1}\right]}

\newcommand{\Cov}[1]{\mbox{Cov}\left[{#1}\right]}

\renewcommand{\eqref}[1]{(\ref{eq:#1})}

\newenvironment{mx}[1]{\left(\begin{array}{#1}}{\end{array}\right)}
{\end{minipage}}

\newcommand{\MSE}{\mbox{MSE}}
\newcommand{\ECD}{\mbox{ECD}}
\newcommand{\WIMSE}{\mbox{WIMSE}}
\newcommand{\KL}{\mbox{KL}}

\newcommand{\X}{\mathbf{X}}

\newcommand{\Z}{\mathcal{G}}

\newcommand{\y}{\mathbf{y}}
\newcommand{\N}{\mathcal{N}}
\newcommand{\R}{\mathbb{R}}
\newcommand{\Rr}{\mathbf{R}}

\newcommand{\argmax}[2]{\smash{\mathop{{\rm argmax}} \limits_{#1}}\,#2}
\newcommand{\argmin}[2]{\smash{\mathop{{\rm argmin}} \limits_{#1}}\,#2}

\begin{document}
\title{An adaptive kriging method for solving nonlinear inverse statistical problems}

\author{Shuai Fu\thanks{EDF Lab Chatou \& Universit\'e Paris XI}, Mathieu Couplet\thanks{EDF Lab Chatou} and Nicolas Bousquet\thanks{EDF Lab Chatou \& Institut de Math\'ematique de Toulouse: nicolas.bousquet@edf.fr }}




\maketitle

\begin{abstract}
 {In various industrial contexts, estimating the distribution of 
unobserved random vectors $X_i$ from some noisy indirect observations
$H(X_i)+U_i$ is required. If the relation between
$X_i$ and the quantity $H(X_i)$, measured with the error $U_i$,
is implemented by a CPU-consuming computer model $H$,
a major practical difficulty is to perform the statistical
inference with a relatively small number of runs of $H$.}
Following \citet{fu14}, a Bayesian statistical framework is
considered \textcolor{black}{to make use of possible prior knowledge on the parameters
of the distribution of the  {$X_i$}, which is assumed Gaussian.} Moreover, a 
\textcolor{black}{Markov Chain Monte Carlo} (MCMC) algorithm is carried out to estimate their
posterior distribution by replacing $H$ by a kriging metamodel
build from a limited number of \textcolor{black}{simulated experiments}.
Two heuristics, involving two different criteria to be optimized, are proposed
 to sequentially design these computer experiments in the limits
of a given computational budget. The first criterion is a Weighted Integrated
Mean Square Error (WIMSE) \citep{picheny10}. The second one, called
Expected Conditional Divergence (ECD),  {developed in the spirit of the}
Stepwise Uncertainty Reduction (SUR)  {criterion} \citep{vazquez09,bect12}, is based on the  {discrepancy} between two consecutive approximations of the target posterior distribution. 
Several numerical comparisons conducted over a toy example then a motivating real case-study
show that such adaptive designs can significantly outperform the classical choice of a 
maximin Latin Hypercube Design (LHD) of experiments. \textcolor{black}{Dealing with a major concern in hydraulic engineering, a particular emphasis is placed upon the prior elicitation of the case-study, highlighting the overall feasibility of the methodology.}  
Faster convergences and manageability considerations lead to
recommend the use of the ECD criterion in practical applications. 
\end{abstract}

{\small 
\textbf{Keywords:}
{Inverse statistical problems};
 {Bayesian inference};
 {Kriging};
 {Adaptive design of experiments};
 {Metropolis-Hasting-within-Gibbs algorithm};
 {Prior elicitation}.
 }


\section{Introduction}
\label{intro}

In many industrial problems, engineers have to deal with uncertain quantities
which cannot be directly measured. Moreover, \textcolor{black}{some of them can suffer} from
some inherent variability.
For instance, in hydraulics, the assessment of a risk of flooding usually
depends on some quantities, called coefficients of Manning-Strickler, which
represent the roughness of the river bed. Because rivers are complex changeable
systems, it appears reasonable to consider these coefficients as
random variables. Although they cannot be directly measured,
it appears possible to estimate their randomness from flooding data
by means of computer simulation.

Estimating the probability distribution of such random unobserved variables
involves some observations  {$Y_i\in\R^p$} (e.g., water levels),
$i=1\ldots{}n$, and a computer model $H$ (e.g., Saint-Venant equation solver)
which links the unobserved variables of interest  {$X_i\in\R^q$}
(e.g., Manning-Strickler coefficients) to the $Y_i$:
\begin{eqnarray}\label{eq:model}
Y_i = H(X_i, d_i) + U_i
\end{eqnarray}
where $d_i\in\R^{q_2}$ stands for some known or observed quantities
and where $U_i$ represents some unobserved measurement errors.
A Gaussian framework is adopted in this article:
the $\trans{(X_i \; U_i)}$ are assumed to be independent Gaussian random vectors
such that
\begin{eqnarray}\label{eq:gaussianFramework}
\vr{X_i\\ U_i} \sim \mathcal{N}_{q+p}(\vr{m\\ 0}, \begin{mx}{cc}
C & 0\\
0 & R
\end{mx})
\end{eqnarray}
where $\mathcal{N}_k(\mu, \Sigma)$ is the $k$-dimensional Gaussian distribution
of mean $\mu$ and covariance matrix $\Sigma$.
The issue is then to estimate the unknown parameters $\theta=(m, C)$ of the
probability distribution of the $X_i$ from some field data $(y_i, d_i)$\footnote{
 {Where $y_i$ is a realization of the random vector $Y_i$.}},
$1\leq i\leq n$, given $H$ and the error covariances $R$.
The accuracy of measurements is generally given or can be assessed:
the assumption that $R$ is known is a sound basis for the inference, since
it prevents from a problem of non-identifiability of ($\theta, R$).

From the general perspective of the analysis of some independent measurements
$y_i$ performed on similar systems (under conditions $d_i$),
this statistical model enables to capture the inherent variability of some
variables $X_i$ in the population which is studied.
For instance, mechanical tests generally involve a production-lot
population of components whose precise characteristics (\textit{e.g.}
Young's Modulus or thermal expansion coefficient) suffer from a non negligible
variability.\\

The major practical obstacle to the estimation of $\theta$ is the CPU cost and time 
 needed to evaluate $H(x, d)$, given an input $(x, d)\in\R^Q$ ($Q=q+q_2$).
In hydraulics, one run of $H$ takes typically few hours per CPU.
Several methods have been developed to tackle this difficulty.
\citet{celeux07} considered a maximum likelihood estimation by Expectation-Conditional
Maximisation Either (ECME) \citep{liu94} based on an iterative linearisation of $H$:
this algorithm should be avoided if the nonlinearities of $H$ relative to $x$ are
significant, otherwise it can be very efficient.
\citet{barbillon11} proposed to couple a Stochastic Expectation Maximisation (SEM)
algorithm \citep{celeux88} with a kriging metamodelling of $H$ to improve
the robustness of the estimation.

Kriging, also known as Gaussian Process (GP) regression, was suggested by
\citet{sacks89a} to deal with CPU-expensive computer models.
The purpose of this metamodelling technique is to build an accurate surrogate
model of $H$ from some computer experiments (some runs of $H$).
Then a crucial question is how to determine  the Design of
these Experiments (DoE). Several methods of calibration of computer models
relying on kriging were proposed by \citet{kennedy01} and \citet{bayarri07}. Although their
statistical models are close to the one postulated here,
an important difference is that the $X_i$ are assumed to be random in this article,
whereas the unknown inputs $x$ are part of the parameters $\theta$ to estimate
in their studies.

Hereafter, the Bayesian framework suggested by \citet{fu14}, which involves
a kriging of $H$, is considered. It allows to take account of prior information
 about the $X_i$ (which could possibly arise from expert or past assessments) through the definition of a so-called prior
probability distribution for $\theta$, the density of which being denoted $\pi(\theta)$.
A Metropolis-Hastings-within-Gibbs Markov Chain Monte-Carlo (MCMC) sampling can
then be carried out to estimate the posterior distribution of $\theta$
(given the field data). The benefit of kriging is twofold: extensive sampling
gets feasible, and the uncertainty about $H$ can be accounted for by embedding
the GP into the statistical model and the MCMC procedure.

From a Bayesian point of view, there is
no reason to drop the uncertainty on $H$ by only keeping the kriging predictor
$\hat{H}(.)$.
Besides, the development of purpose-oriented adaptive DoE approaches, such as
the Stepwise Uncertainty Reduction (SUR) \citep{vazquez09,bect12},
is then made possible. Such approaches seek a trade-off between shrinking the
uncertainty on $H$ (which is measured by the kriging covariance) as much as possible,
and exploring the most interesting areas of the input space of $H$ regarding
the considered objective. A classical example comes from the field of global
optimisation where the Expected Improvement criteria was proposed by
\citet{mockus78,jones98}. The purpose of this article is to contribute to the
definition of efficient adaptive DoE algorithms
for solving the inverse statistical problem specified
earlier.\\

The article is organised as follows. Section \ref{review} gives details about
kriging metamodelling, as well as the maximin Latin Hypercube Design (maximin LHD)
which provides us with a first knowledge about the computer model $H$
(before starting a purpose-oriented exploration of the input space of $H$).
In Section \ref{prior_MCMC}, the method used to specify an informative prior
$\pi(\theta)$, then the inference by MCMC, are described. Afterwards,
two methods, called  {Expected Conditional Divergence (ECD)}
and Weighted Integrated Mean Square Error (WIMSE), which derive from two
purpose-oriented criteria to optimise, are proposed to sequentially enrich
the DoE in Section~\ref{ECD} and Section~\ref{WIMSE}.
\textcolor{black}{Numerical studies are conducted on on toy example in Section~\ref{tests} to compare the efficiency of 
these approaches with a posterior approximation standed on a static space-filling design (maximin LHD). Finally, the full
methodology is run through over a real hydraulic computer model:  its input roughness parameters are 
 calibrated from noisy observations of water levels. }
Section~\ref{conclusion} concludes the article by giving major directions for
further work.

\section{Kriging and maximin LHD space-filling design}
\label{review}

This section recall some basics of kriging and of designing computer experiments
which matter for the remainder of the article.

\subsection{Kriging}
\label{sec:kriging}

Kriging is a geostatistical method \citep{matheron71} which was suggested by
\citet{sacks89a} to build a cheap surrogate model of a computer model, from
a limited of runs of the latter, over a hypercube $\Omega\subset\R^Q$.
This method has known a growing interest in
metamodelling with the writings of \citet{koehler96,stein99,kennedy01,santner03},
amongst others. In this section, a scalar function $\dfc{h}{\Omega}{\R}$
is considered: in the case of a vector-valued function $\dfc{H}{\Omega}{\R^p}$,
each component $h_i(.)$ of $H(.)$ can be ``kriged'' independently from the others,
as done in the numerical experiments in Section~\ref{tests}. 

A usual manner to present kriging is starting from the premise that the considered
function $h(.)$ is a particular realization of an underlying GP $\mathcal{H}(.)$:
\begin{eqnarray}\label{eq:krig}
\forall z\in\Omega, & & \mathcal{H}(z) = F(z)\,\beta + \Z(z),
\end{eqnarray}
where $\beta$ is a vector of $\R^K$, where $F(z) = \vr{\!\!f_1(z)\cdots f_K(z)\!\!}$
with $\dfc{f_k}{\R^Q}{\R^p}$, $1\leq k\leq K$, a family of linearly independent
functions, and where $\Z$ is a centered GP ($\Esp{\Z(z)}=0$, for
all $z\in\Omega$). The GP hypothesis means that
$\trans{\vr{\!\!\Z(z_1)\:\cdots\:\Z(z_k)\!\!}}$
is a $k$-dimensional Gaussian vector for any set $\{z_1, \cdots, z_k\}\in\Omega$
and any $k\geq 1$. Although it may appear artificial, this assumption leads to a
flexible statistical model which has been applied successfully in many contexts, and,
in a Bayesian perspective, it can be interpreted as the definition of a prior on $h$ 
\citep{rasmussen06}.
For any $(z,w)\in\Omega^2$, the mean function $\dfc{\mu}{\Omega}{\R}$ of
$\mathcal{H}$ is defined by $\mu(z) = \Esp{\mathcal{H}(z)} = F_z\,\beta$,
and the covariance function $\dfc{K}{\Omega^2}{\R}$ of $\Z$ (and $\mathcal{H}$)
by $\Cov{\Z(z),\Z(w)} = K(z,w)$.
In the following, as most often assumed by authors when modelling computer models,
$\Z$ is stationary, thus $K(z,w)$ only depends on $z-w$:
$K(z,w)=\sigma^2\,K(z-w)$ by abuse of notation, with $K(0)=1$.

Let $D_N=\{z_1, \cdots, z_N\}\subset\Omega$ be a DoE associated to observations
$h_N\in\R^N$,
and $\trans{\mathcal{H}_N}=\vr{\!\!\mathcal{H}(z_1)\cdots\mathcal{H}(z_N)\!\!}$,
then, from a direct application of a classical
theorem relative to the conditioning of Gaussian vectors, the process $\mathcal{H}$
conditioned by the observations,
that is $\mathcal{H}|\mathcal{H}_N\!=\!h_N$, is still a GP over $\Omega$ with mean function $\dfc{\mu_{D_N}}{\Omega}{\R}$ and
covariance function $\dfc{K_{D_N}}{\Omega^2}{\R}$. Namely, for all $(z,w)$,
\begin{eqnarray}
h_N(z)  & \sim & {\cal{N}}\left(\mu_{D_N}(z), {K_{D_N}(z,\cdot)}\right) \label{dist_GP}
\end{eqnarray}
where
\begin{eqnarray}
 & \mu_{D_N}(z) = F(z)\,\beta + K_{z,N} {K_{N,N}}^{-1} (h_N\!-\!F_N\,\beta)
\label{eq:mu}\\[1ex]
 & K_{D_N}(z,w) = K(z,w) - K_{z,N}{K_{N,N}}^{-1}\trans{K_{w,N}} 
\end{eqnarray}
with $K_{z,N}=\vr{\!\!K(z,z_1)\cdots K(z,z_N)\!\!}$ (idem for $K_{w,N}$) and
${[K_{N,N}]}_{i,j} = K(z_i, z_j)$.
Of course, $K_{D_N}(z,z) < K(z,z)$: the more observations are available,
the less uncertainty on $\mathcal{H}$ remains.
If $K(.,.)$ and $\beta$ are known, then $\mu_{D_N}(z)$ is the
kriging predictor of $\mathcal{H}(z)|\mathcal{H}_N\!=\!h_N$, that is its
Best Linear Unbiaised Predictor (BLUP), and $K_{D_N}(z,w)$ is the kriging
covariance, that is the covariance function
of the error of prediction. In particular, $K_{D_N}(z,z)$ is the Mean Square
Error (MSE) of the BLUP at $z$.

Furthermore, if $K(.,.)$ is known but $\beta$ unknown (universal kriging),
then the generalised least-square estimator
\begin{eqnarray}\label{eq:beta}
\hat{\beta} = {(\trans{F_N}K_{N,N}F_N)}^{-1}\trans{F_N}K_{N,N}h_N
\end{eqnarray}
of $\beta$ is also the maximum likelihood estimator, and the BLUP
$\hat{h}_{D_N}(z)$ of $\mathcal{H}(z)|\mathcal{H}_N\!=\!h_N$ is obtained
by substituting $\beta$ by $\hat{\beta}$ in Equation~\eqref{mu}.
Last but not least, the kriging covariance which is associated to
$\hat{h}_{D_N}(z)$ is then
\begin{eqnarray}
K_{D_N}(z,w) & = & K(z,w) - K_{z,N}{K_{N,N}}^{-1}\trans{K_{w,N}} \nonumber \\
 & + & \trans{(F(z) - \trans{F_N}{K_{N,N}}^{-1}\trans{K_{z,N}})} \label{eq:krigingCov} 
{(\trans{F_N}{K_{N,N}}^{-1}F_N)}^{-1} \\
& & \ \ \times \ 
(F(w) - \trans{F_N}{K_{N,N}}^{-1}\trans{K_{w,N}}) \nonumber
\end{eqnarray}
(we use the same notation as before - $\beta$ known - for the sake of simplicity)
with $[{F_N}]_{i,j} = f_j(z_i)$.
It can be seen as an approximation of the covariance function of the GP
$\mathcal{H}|\mathcal{H}_N\!=\!h_N$ and, together with $\hat{h}_{D_N}(z)$,
is generally used, for example in \citep{bect12}, to model the uncertainty on the
Gaussian vector
$\trans{\vr{\!\!\mathcal{H}(w_1)\cdots\mathcal{H}(w_M)}}|\mathcal{H}_N\!=\!h_N$
for any set $\{w_1,\cdots,w_M\}\subset\Omega$; see also \citet{santner03,bachoc13} for other precisions.
Following \citet{fu14}, $K_{D_N}(z,w)$ is used in the MCMC procedure 
to account for the dependence between the missing data $X_i$ due to
the uncertainty on (each component $h_i(.)$ of)
the computer model $H(.)$; see Section~\ref{prior_MCMC} for more details.
The induced Mean Square Error $MSE_{D_N}:z\mapsto K_{D_N}(z,z)$
 {plays} an important role in Section~\ref{WIMSE}.

In practical applications, $K(.,.)$ is unknown and is estimated, thanks
to a model $K_\psi(.,.)$ parametrised by $\psi\in\R^L$, by different
techniques such as maximum likelihood (as hereafter) or cross-validation.
In the remainder of this article, the plug-in estimates obtained by
replacing $K(.,.)$ by $K_{\hat{\psi}}(.,.)$, with $\hat{\psi}$ the estimator
of $\psi$, are employed.

\subsection{Design of experiments ({\it maximin}-Latin Hypercubic Designs)}

Obviously, the predicting accuracy of kriging highly depends on the DoE $D_N$. 
Following \citet{picheny10}, it is possible to distinguish three kinds of DoEs:
\begin{itemize}
\item {\it space-filling} designs, which aim to fill the input space with a finite number of points independently of the considered model (e.g., {\it maximin}-LHD); 
\item {\it model-oriented} designs, which attempt to build a suited DoE accounting for the features of the model $H$ or the metamodel
 {(e.g. IMSE, see Section~\ref{WIMSE:intro})};
\item {\it purpose-oriented} designs, which account for the final aim of the study to find the best adapted DoE (e.g., to compute an exceedance probability by accelerated Monte Carlo methods).
\end{itemize}

 {In this article, a {\it purpose-oriented} DoE is built in an adaptive way.
A first calibration of the covariance parameters is  {performed} from an initial
{\it maximin}-LHD, then the DoE is sequentially improved using sequential strategies,
which are detailed in sections~\ref{ECD} and~\ref{WIMSE}.
The concept of LHDs was introduced in \cite{mckay79}; such designs ensure a good
coverage of the interval to which each scalar variable belongs.
Then \cite{johnson90} proposed the {\it maximin} distance criterion to optimize LHDs.
{\it Maximin} means maximizing the minimum inter-site distance between the set of $N$ points:
\begin{eqnarray*} 
\delta_D & = & \min_{i\neq j} \Vert z_{(i)}-z_{(j)}\Vert_2.
\end{eqnarray*}
Therefore, the {\it maximin} criterion prevents the points of the design to be close
to each other. In the present work, {\it maximin}-LHDs are obtained by the
algorithm of \citet{morris95}.}


\section{Bayesian statement and inference}
\label{prior_MCMC}

\subsection{Prior elicitation}\label{prior.elicitation}

In the Bayesian statistical framework favored in \cite{fu14}, a Gaussian-Inverse Wishart prior distribution was elicited:
\begin{eqnarray}
m\,|\,C & \sim & \N_q(\mu,C/a),\label{eq:InfoPrior1}\\
C & \sim & \mathcal{IW}_q(\Lambda,\nu).\label{eq:InfoPrior2}
\end{eqnarray}
\textcolor{black}{
This prior can be assimilated to the posterior distribution of virtual data given a noninformative prior, which presents some advantages in subjective Bayesian analysis \citep{Bousquet2015}. Especially, a clear sense can be given to hyperparameters $(\mu,a,\Lambda,\nu)$, which simplifies prior calibration.} 

\textcolor{black}{
Indeed, $a$ can be understood as the size of {\it virtual} sample of data $X$, that modulates the strength of the practicioner's belief in prior information (for instance provided by subjective experts). It should be calibrated under the constraint $a<n$ to ensure that the posterior behavior is mainly driven by objective data information. A default (let say, ``objective") choice is $a=1$.} 

\textcolor{black}{Furthermore, $\mu$ is the prior predictive mean, median and most probable value of $X$, which can be estimated by a measure of central tendency provided by past calibration results in close situations. In the motivating case-study explored in Section \ref{numerical.experiments}, such information was found by bibliographical researches (Table \ref{prior.strickler}).} 

\textcolor{black}{Finally, denoting $\mathbf{X,Y}$ the set of missing and truly observed data, the reparametrizations 
$\Lambda  =  (a+1)\cdot C_{e}$ and $\nu  =  a+q+2$
imply that the conditional posterior distribution of $C$ given $m$ is the Inverse Wishart distribution 
$\mathcal{IW}\Big((a+1)\,C_{e}+(n+1)\,\hat{C}_n, \,\nu+n+1\Big)$
with $\hat{C}_n  =  \frac{1}{n} \sum_{i=1}^n (m-x_i)(m-x_i)^T$, the  {expectation} of which being
\begin{eqnarray*}
\mathbb{E}[C\,|\,m,\mathbf{X,Y}] & = & \frac{a+1}{a+n+2}\cdot C_{e} + \frac{n+1}{a+n+2}
\cdot \hat{C}_n. 
\end{eqnarray*}
This last expression highlights the meaning and influence of $a$ as a virtual size. The components of $C_{e}$ are to be calibrated in function of prior knowledge on $X$ too, expressed through its predictive prior distribution, which is a decentered Student law:
\begin{eqnarray*}
X & \sim & \mbox{St}_q\Big(\mu,\,\frac{(a+1)^2}{a(a+3)}C_{e},\,a+3 \Big)
\label{chap6:predic}
\end{eqnarray*}
with mean vector $\mu$ and covariance matrix $\frac{a+1}{a} C_{e}$.
Again, in the case-study that motivated this work, prior information on the ratio between average values and standard deviation of Strickler-Manning coefficients was available (Figure \ref{usa9600}), which allowed for a full prior calibration (Section \ref{numerical.experiments}). 
}

\subsection{Posterior computation}\label{posterior.computation}

  A Gibbs sampler \citep{tierney96} was proposed to compute the posterior distribution of  $\theta=(m,C)$. Actually, replacing the expensive-to-compute function $H$ with a kriging emulator $\widehat{H}$, as in \citet{barbillon11}, and  introducing a new {\it emulator error} $\MSE$, the Gibbs sampler can be adapted as follows:


\paragraph{Gibbs sampler (at the $(r+1)$-th iteration)}\label{algo:Gibbs} 
\rule[0.5ex]{6cm}{0.1mm}
\vspace{0.1cm}
\small

\texttt{Given $(m^{(r)}, C^{(r)}, \mathbf{X}^{(r)})$ for $r=0,1,2,\dots$, generate:} \\

\begin{enumerate}
\item $C^{(r+1)}|\dots \sim \mathcal{IW}\Big(\Lambda+\sum_{i=1}^n (m^{(r)}-X_i^{(r)})(m^{(r)}-X_i^{(r)})'+a(m^{(r)}-\mu)(m^{(r)}-\mu)', \,\nu+n+1\Big)$,
\item $m^{(r+1)}|\dots \sim \mathcal{N}\Big(\frac{a}{n+a}\mu+\frac{n}{n+a}\overline{\mathbf{X}_n^{(r)}},\, \frac{C^{(r+1)}}{n+a}\Big)$
 where
$\overline{\mathbf{X}_n^{(r)}}=n^{-1} \sum_{i=1}^n X_i^{(r)}$, 

\item $\mathbf{X}^{(r+1)}|\dots  \propto |\mathbf{R}+\mbox{MSE}^{(r+1)}|^{-\frac{1}{2}}\cdot \exp\Bigg\{-\frac{1}{2} \sum_{i=1}^n  (X_i^{(r+1)}-m^{(r+1)})'\Big[C^{(r+1)}\Big]^{-1}(X_i^{(r+1)}-m^{(r+1)}) - \frac{1}{2}\Big(\Big(\mathcal{Y}_1-\widehat{H}_{N,1}^{(r+1)}\Big)',\dots,\Big(\mathcal{Y}_n-\widehat{H}_{N,n}^{(r+1)}\Big)'\Big)
\Big(\mathbf{R}+\mbox{MSE}^{(r+1)}\Big)^{-1}
\left( \begin{array}{c} \mathcal{Y}_1-\widehat{H}_{N,1}^{(r+1)}\\
\vdots \\ \mathcal{Y}_n-\widehat{H}_{N,n}^{(r+1)} \end{array} \right)\Bigg\}\\
\label{eq:posterX}$

\texttt{where $\widehat{H}_{N,i}^{(r+1)}=\widehat{H}_N(X_i^{(r+1)},d_i)$ and
 $\mbox{MSE}^{(r+1)}=\mbox{MSE}(\mathbf{X}^{(r+1)},\mathbf{d})$ is the block diagonal matrix } 
 \begin{eqnarray*}
 \mathbf{\mbox{MSE}}(\mathbf{X}^{(r+1)},\mathbf{d}) & = & \begin{array}{cl}
 \left(\begin{array}{ccc} 
\mbox{MSE}_1(\mathbf{X}^{(r+1)},\mathbf{d}) & & {\bf 0}\\ 
 & \ddots & \\ 
  {\bf 0} & &\mbox{MSE}_p(\mathbf{X}^{(r+1)},\mathbf{d})
  \end{array}\right)\,\,
   & \begin{array}{ll}
\left. \begin{array}{l} \\  \end{array} \right\rbrace & n \textrm{ lines} \\
\\
\left. \begin{array}{l} \\\end{array} \right\rbrace & n \textrm{ lines} \\
\end{array}
\end{array}
\end{eqnarray*} 
\end{enumerate}
\rule[0.5ex]{\textwidth}{0.1mm}

\normalsize

\vspace{0.5cm}
\noindent In the third step, the variance matrices  $\mbox{MSE}_j(\mathbf{X}^{(r+1)},\mathbf{d})\in\mathcal{M}^{n\times n}$ are defined by
\begin{eqnarray*}
\mbox{MSE}_j(\mathbf{X}^{(r+1)},\mathbf{d}) & = & \mathbb{E}\left(\left(\mathcal{H}_j(\mathbf{X}^{(r+1)},\mathbf{d})-\widehat{H}_j(\mathbf{X}^{(r+1)},\mathbf{d})\right)^2\,|\,\mathbf{H }_{D_N}\right),
\end{eqnarray*}
for $j=1,\ldots,p$, where $\mathcal{H}_j$ denotes the $j$-th dimension of the Gaussian process $\mathcal{H}$. Moreover, 
 \begin{eqnarray*}
 \Rr & = & \begin{array}{cl}
 \left(\begin{array}{ccc} 
\Rr_{1} & & {\bf 0}\\ 
 & \ddots & \\ 
  {\bf 0} & &\Rr_{p}
  \end{array}\right)\,\,
   & \begin{array}{ll}
\left. \begin{array}{l} \\  \end{array} \right\rbrace & n \textrm{ lines} \\
\\
\left. \begin{array}{l} \\\end{array} \right\rbrace & n \textrm{ lines} \\
\end{array}
\end{array},  \textrm{ with } \, \Rr_{i} \, = \,  \left(\begin{array}{ccc}
 R_{ii} & & 0 \\
 & \ddots & \\
 0 & & R_{ii}\end{array}\right),
\end{eqnarray*} 
where $R_{ii}$ is the $i-$th diagonal component of the diagonal variance matrix $R$. It is worth noting that this third conditional distribution does not belong to any closed form family of distributions. Therefore a Metropolis-Hastings (MH) step is used to simulate $\mathbf{X}^{(r+1)}$  (see Appendix A).
 
As discussed in \cite{fu14}, the use of the MCMC algorithms involves many possible errors. According to experimental trials, the accuracy of the metamodel plays a critical role in the the estimation problem. MCMC algorithms can produce Markov chains converging towards the desired posterior distribution. However, if the function $H$ is really badly approximated, apart from the {\it algorithmic error} introduced by the MCMC algorithm, the result can also suffer from an {\it emulator error}.

\section{The Expected Conditional Divergence criterion for adaptive designs}
\label{ECD}

The two following sections address the issue of building adaptive designs of experiments, by proposing two strategies. In this section, a criterion called $\ECD$ (Expected Conditional Divergence) is  {built}, which  {can be seen as an adaptation of} the Expected Improvement criterion proposed in \citet{jones98}.
 {Let us notice that the expected divergence criterion proposed in the next section,
although close to a Stepwise Uncertainty Reduction (SUR) criterion, does not derive
from the SUR formulation of \citet{vazquez09,bect12}. The latter would lead to a
more challenging approach from a computational perspective in our context.}

\subsection{Principle}

Ideally, the posterior distribution of the parameters $\theta=(m,C)$ after adding a new point $z_{(N+1)}$ to the current DoE $D_N$ should be as close as possible to the posterior distribution knowing the original function $H$, i.e. a relevant discrepancy measure between the two relative distributions must be minimized. Based on information-theoretical arguments given in \citet{cover06}, the Kullback-Leibler  {(KL)} divergence 
\begin{eqnarray}
& & \KL\Big(\pi(\theta|{\bf y,d},H)\, ||\,\pi(\theta|{\bf y,d},\mathbf{H}_{D_N}\cup \{H(z)\}))\Big),
\label{eq:idealKL}
\end{eqnarray}
is a good choice of discrepancy measure. Remind that given two densities $p(x)$ and $q(x)$ defined over the same space $\mathcal{X}$, 
\begin{eqnarray*}
\KL(p||q) & = & \int_{\mathcal{X}} p(x)\log\frac{p(x)}{q(x)}\, dx.
\end{eqnarray*}  
Ideally,  the next point $z_{(N+1)}$ should be searched within the feasible region $\Omega$, as the global minimum of this divergence. 
\textcolor{black}{But obviously, the unknown term $\pi(\theta|{\bf y,d},H)$ makes this formulation intractable. But  a tractable sub-optimal criterion can be heuristically derived from it by the following rationale.}  It must be noticed that
\begin{eqnarray}
z_{(N+1)} & = & \argmin {z\in\Omega}\, \KL\Big(\pi(\theta|{\bf y,d},H)\, ||\, \pi(\theta|{\bf y,d},\mathbf{H}_{D_N}\cup \{H(z)\})\Big),\nonumber\\
& = & \argmin {z\in\Omega}\, \KL\Big(\pi(\theta|{\bf y,d},H)\, ||\, \pi(\theta|{\bf y,d},\mathbf{H}_{D_N}\cup \{H(z)\})\Big) \nonumber\\
& & \ \ \ - \ \KL\Big(\pi(\theta|{\bf y,d},H)\, ||\, \pi(\theta|{\bf y,d},\mathbf{H}_{D_N})\Big),\nonumber\\
& = & \argmax {z\in\Omega}\, \int_{\theta \in \Omega} \pi(\theta|{\bf y,d},H)\, \log\frac{\pi(\theta|{\bf y,d},\mathbf{H}_{D_N}\cup \{H(z)\})}{\pi(\theta|{\bf y,d},\mathbf{H}_{D_N})}\, d \theta\nonumber.
\end{eqnarray}
The  {intractable} target density $\pi(\theta|{\bf y,d},H)$ has to be replaced with its best available approximation, which is $\pi(\theta|{\bf y,d},\mathbf{H}_{D_N}\cup \{H(z)\})$. Under the kriging assumptions,  for any $z$ this distribution is closer of $\pi(\theta|{\bf y,d},H)$ than $\pi(\theta|{\bf y,d},\mathbf{H}_{D_N})$. Therefore, a sub-optimal version of the idealistic criterion is:
\begin{eqnarray*}
z_{(N+1)} & = &  \argmax {z\in\Omega}\, \KL\Big(\pi(\theta|{\bf y,d},\mathbf{H}_{D_N}\cup \{H(z)\})\, ||\,\pi(\theta|{\bf y,d}, \mathbf{H}_{D_N})\Big).\label{eq:KL1}
\end{eqnarray*}
In other words, the chosen  strategy aims at finding the optimal point $z_{(N+1)}$ which modifies the actual distribution $\pi(\theta|{\bf y,d},\mathbf{H}_{D_N})$ as much as possible in an information-theoretic sense.  First proposed by \citet{stein1964} as a loss function, the dissymetric  {KL} divergence between the two consecutive posterior distributions, which   is invariant under one-to-one transformation of the random vector $\theta$, has an operative interpretation as the loss of information (in natural information units or {\it nits}) which may be expected by choosing the baddest approximation $\pi(\theta|{\bf y,d}, \mathbf{H}_{D_N})$ instead of the best (available) $\pi(\theta|{\bf y,d},\mathbf{H}_{D_N}\cup \{H(z)\})$ \citep{cover06,berger2015}.  \\ 

The preceding formulation is not satisfactory yet, since one evaluation of the criterion requires one evaluation of $H$, which is time-consuming. However, in the spirit of $\mbox{EGO}$, it is possible to derive a new criterion considering the following Gaussian process based on the available observations $\mathbf{H}_{D_N}$ instead of $H$:
\begin{eqnarray}\label{eq:hNz}
h_N(z) & := & \mathcal{H}(z)\,|\,\mathbf{H}_{D_N},
\end{eqnarray}
which follows the normal distribution given in (\ref{dist_GP}). Thus, we define the {\it expected divergence criterion}:
\begin{eqnarray}
z_{(N+1)} & = &  \argmax {z\in\Omega} \mathbb{E}_{\pi(h_N)}\left[ \KL\left(\pi(\theta|{\bf y,d},\mathbf{H}_{D_N}\cup \{h_N(z)\})\, \right.\right. \label{eq:expKL}  \\
& & \hspace{3.5cm}  \left.\left. ||\,\pi(\theta|{\bf y,d}, \mathbf{H}_{D_N})\right)\right].\hspace*{1.5cm}
\nonumber
\end{eqnarray}
The idea of considering the Gaussian variable $h_N(z)$ rather than the predictor $\widehat{H}_N(z)$ allows  to account for the uncertainty introduced by the kriging methodology, while it requires usual Monte Carlo methods to approximate the double integrals, i.e. the expectation and the  {KL divergence}. 

 {Even if no run of $H$ is required, the evaluation of this expected divergence
criterion requires many calculations. In the next section, a heuristic is proposed to
shrink the computational cost of the approach.}


\subsection{The Expected Conditional Divergence  {heuristic}}
\label{CD:algo}

Preliminary experiments showed that the criterion defined in \eqref{expKL} is
 {generally too expensive to be useful, except for extremely CPU-consuming code $H$.
The main reason  is that any test of a new point $z$ requires to run a Gibbs sampler.}
Therefore a last adaptation of the criterion is proposed: the Expected Conditional Divergence  ($\ECD$) criterion depends only on the intermediate full-conditional posterior distributions of $\theta$. More precisely, at the $(r+1)$-th iteration of the Metropolis-Hastings-within-Gibbs algorithm, the strategy is defined as:
\begin{eqnarray}\label{eq:critSequ}
z_{(N+1)} & = & \argmax {z\in\Omega} \ECD(z)
\end{eqnarray}
with
\begin{eqnarray}\label{eq:ecd}
\ECD(z) & = & \mathbb{E}_{\pi(h_{N})}\left[ \KL\Big(\pi(\theta|\tilde{\X}^{(r+1)}(z))\, ||\,\pi(\theta|\X^{(r+1)})\Big)\right], \label{ecdtheo}
\end{eqnarray}
where $\X^{(r+1)}$ and $\tilde{\X}^{(r+1)}(z)$ denote the missing data samples simulated from
\begin{eqnarray*}
\X^{(r+1)} & \sim & \pi\left(\cdot|{\bf y,d},\theta^{(r+1)},\mathbf{H}_{D_N}\right),\\
\tilde{\X}^{(r+1)}(z) & \sim & \pi\left(\cdot|{\bf y,d},\theta^{(r+1)},\mathbf{H}_{D_N}\cup \{h_{N}(z)\}\right).
\end{eqnarray*}
It is worth noting that in the $\ECD$ criterion, the final posterior distribution of
$\theta$ is replaced by its sequential conditional posterior distribution at the
$(r+1)$-th iteration. 
{At the $(r+1)$-th iteration of the Gibbs sampling, given a candidate
$z$ to enrich the DoE, this heuristic enables to compute a value $\ECD(z)$ which is
likely a sufficient approximation of the expected divergence criterion for the global
algorithm to perform well.
Moreover, once $\ECD(z)$ has been evaluated, the computation of $\ECD(z')$ at a new
candidate $z'$ takes benefit of the computations performed during the
calculation of $\ECD(z)$ (sampling of $\X^{(r+1)}$ by Metropolis-Hastings, then
sampling of $\theta$ given $\X^{(r+1)}$) and does not require a full Gibbs sampling
anymore (just the MCMC sampling of $\tilde{\X}^{(r+1)}(z)$, then the sampling of
$\theta$ given $\tilde{\X}^{(r+1)}(z)$). Hence it allows an exploration
of the input space (optimization of $\ECD$) for a acceptable CPU-cost.

Finally, using a standard Monte-Carlo estimator to estimate the
expectation of the KL divergence according to $\pi(h_{N})$ (see~\eqref{ecd}),
the ECD heuristic algorithm proceeds as follows:


\paragraph{ECD strategy} 
\rule[0.5ex]{11cm}{0.1mm}
\vspace*{0.1cm}
\small
\texttt{~\\[-2ex]Given $(m^{(0)}, C^{(0)}, \mathbf{X}^{(0)})$, an initial
design $D_N$ with the corresponding evaluations $\mathbf{H}_{D_N}$ of $H$:
\begin{enumerate}
 \item $r := 0$.
 \item Perform $k$ new Gibbs iterations (Section~\ref{posterior.computation});
$r := r+k$: this gives $\theta^{(r+1)}$.
 \item Sample $\X^{(r+1)}$ from $\pi\left(\cdot|{\bf y,d},\theta^{(r+1)},\mathbf{H}_{D_N}\right)$ (see Appendix~A).
 \item Sample $\Upsilon=\{\theta_1,\dots,\theta_{L_2}\}$ from $\pi(\cdot|\X^{(r+1)},{\bf y,d})$
(explicit distribution: see steps~1 and~2 of Section~\ref{posterior.computation}).
 \item Get a new point $z_{(N+1)}$ to enrich the DoE by the optimization of $\ECD$
(simulated annealing, see Appendix~C): for any $z$, assess $\ECD(z)$ if needed by:
 \begin{enumerate}
  \item Generate $M$ samples $(h^1_N(z),\dots,h^M_N(z))$ according to \eqref{hNz}
and build $M$ corresponding emulators $(\widehat{H}^1_{N+1}(z),\dots,\widehat{H}^M_{N+1}(z))$ with $\widehat{H}^i_{N+1}(z)$ based on the dataset $\mathbf{H}_{D_N}\cup \{h^i_N(z)\}$
(no re-estimation of the covariance function parameters $\psi$, see Section~\ref{sec:kriging}).
  \item for $1\leq i\leq M$, 
   \begin{description}
    \item[(i)] Sample $\tilde{\X}^{(r+1),i}(z)$ from $\pi(\cdot|{\bf y,d},\theta^{(r+1)},\widehat{H}^i_{N+1}(z))$ (see Appendix~A).
    \item[(ii)] Sample $\Theta^i=\{\theta^i_1,\dots,\theta^i_{L_1}\}$ with $\theta=(m_1,\dots,m_q,C_{11},\dots,C_{qq})$ from $\pi(\cdot|\tilde{\X}^{(r+1),i}(z),{\bf y,d})$(explicit distribution: see steps~1 and~2 of Section~\ref{posterior.computation}).
   \end{description}
   \item $ECD(z) := \frac{1}{M}\sum_{i=1}^M \widehat{\KL}\Big(\Theta^i\, ||\,\Upsilon\Big)$
where $\widehat{\KL}(.||.)$ denotes the KL divergence estimate (see Appendix~B).
 \end{enumerate}
   \item $D_N := D_N\cup\{z_{N+1}\}$  and
$\mathbf{H}_{D_N} := \mathbf{H}_{D_N}\cup\{H(z_{N+1})\}$ (new run of $H$).
   \item Return to 2 if $\#\mathbf{H}_{D_N}$ is less than the maximal number of runs of $H$.
\end{enumerate}
\rule[0.5ex]{\textwidth}{0.1mm}
}

\normalsize
In our numerical experiments, the optimization (step~5) and the KL divergence
estimation (step 5.(c)) are respectively performed using the simulated annealing (SA)
method \citep{kirkpatrick83} and the Nearest-Neighbor (NN) method of \citet{wang09}
(see Appendices~C and~B for detail): other choices are possible.\\

Let us remark that it can be reasonable to decrease the CPU-cost of ECD by
neglecting the dependencies between the components of $\theta$:
eventually, assuming that these components are independent substantially
decreases the cost of the k-NN KL divergence estimation, since
the multivariate KL divergence is then the sum of univariate KL divergences.
It would be also feasible to suppose that $\theta$ is made up with
independent random vectors (e.g. assuming independence between $m$ and $C$). 
In fact, this technique could be directly applied to the expected divergence criterion
(previous section), thus offers an alternative to ECD. However, it
is not investigated hereafter, because ECD alone leads to a satisfactory trade-off
between efficiency of the DoE enrichment and computational cost, in our
industrial context.}


\section{The Weighted-IMSE criterion for adaptive designs}
\label{WIMSE}

This section is devoted to propose an alternative criterion of adaptive design, by adapting the popular weighted-IMSE criterion \citep{sacks89a,picheny10}, reminded hereinafter, to the Bayesian context of probabilistic inversion. 
 
\subsection{The Integrated MSE criterion}
\label{WIMSE:intro}

The Integrated Mean Square Error (IMSE) criterion \citep{sacks89a} is a measure of the average accuracy of the kriging metamodel over the domain $\Omega$:
\begin{eqnarray*}
\mbox{IMSE}(\Omega) & = &  \int_{\Omega} \MSE(z) \, dz,
\end{eqnarray*}
where $\MSE(z)$ is defined in the Gibbs sampler in $\S$ \ref{posterior.computation}.  
 Given a current design $D_N$ of $N$ points, \citet{picheny10} proposed the following $\WIMSE$ criterion as an alternative approach to improve the prediction accuracy in regions of main interest:
\begin{eqnarray}
\WIMSE(z^*) & = & \int_{\Omega} \MSE\left(z|D_N \cup\{z^*\}\right) w\left(z|D_N,\mathbf{H}_{D_N}\right) \, dz,
\label{eq:wimse1}
\end{eqnarray}
where $\MSE\left(z|D_N \cup\{z^*\}\right)$ denotes the prediction variance by adding the point $z^*=(x^*,d^*)$ into $D_N$ and $w\left(z|D_N,\mathbf{H}_{D_N}\right)$ is a weight function emphasizing the $\MSE$ term over these regions of interest.  The calculation of $\MSE$ does not depend on the expensive evaluation $H(z^*)$ and the weight factor $w$ only depends on the available observations $\mathbf{H}_{D_N}$. The next point to add to the DoE is thus defined by
\begin{eqnarray*}
z_{(N+1)} & = & \arg \min_{z\in\Omega} \WIMSE(z).
\end{eqnarray*}

\subsection{Adaptation to the Bayesian inversion context}
\label{WIMSE:adapt}

Defining the regions of interest is the essential task in applying the $\WIMSE$ criterion. As presented in previous sections, a probabilistic solution to inverse problems is to approximate the posterior distribution of the parameters $\theta=(m,C)$ using a Metropolis-Hastings-within-Gibbs algorithm (cf.  {Section~\ref{posterior.computation}}). Assuming that the $(N+1)-$th new point is added at the $(r+1)-$th iteration of the Gibbs sampling, the weight function is defined by the following formula:
\begin{eqnarray}
w\left(z|D_N,\mathbf{H}_{D_N}\right) & \propto & \prod_{i=1}^n \pi\left(x,d|y_i,\theta^{(r+1)},D_N,\mathbf{H}_{D_N} \right), \label{eq:omega} \\
& \propto &  \prod_{i=1}^n  |\mathbf{R}+\mbox{MSE}(x,d)|^{-\frac{1}{2}} \cdot  \exp\Bigg\{- \frac{1}{2} \Delta_i \Bigg\}\nonumber 
\end{eqnarray}
where 
\begin{eqnarray*}
\Delta_i & =&  (x-m^{(r+1)})'\Big[C^{(r+1)}\Big]^{-1}(x-m^{(r+1)}) \\
& & \  - \ \Big(y_i-\widehat{H}(x,d)\Big)'
\Big(\mathbf{R}+\mbox{MSE}(x,d)\Big)^{-1}
\Big(y_i-\widehat{H}(x,d)\Big),
\nonumber
\end{eqnarray*}
which is derived from the full conditional posterior distribution of $\X$ described in  {Section~\ref{posterior.computation}}. 
It can be considered as a measure of the posterior prediction error. 
 The advantage of this choice is twofold. First, this weight function $\omega$ indicates a potential position for the missing-data $\X$ where the accuracy of the metamodel should be improved. Second, this weight function depends on the observation sample $\y=\{y_1,\dots,y_n\}$, coherently with the Bayesian conditioning process and providing a {\it purpose-oriented} sense to the design. \\

Besides, since the two terms $\MSE(\cdots)$ and $w(\cdots)$ of \eqref{wimse1}
are different in nature, a tuning parameter $\alpha$ is introduced
(as an exponent) to allow for a trade-off between the two.
Therefore the following version of the $\WIMSE$ criterion is proposed:
\begin{eqnarray}
\WIMSE(z^*) & = & \int_{\Omega} \MSE^{\alpha}\left(z|D_N \cup\{z^*\}\right) \, w^{1-\alpha}\left(z|D_N,\mathbf{H}_{D_N}\right) \, dz.
\label{eq:wimse2}
\end{eqnarray}
In this equation, $\alpha$ varying between 0 and 1 makes the criterion more flexible: if $\alpha$ is close to 1, the impact of the weight parameter $\omega$ disappears and the criterion becomes IMSE; if $\alpha$ approaches to 0, the prediction error $\MSE$ will not be accounted for. Experimental trails proved that the choice of $\alpha$ is critical. Furthermore, such a chosen weight function $w$, defined as the product of $n$ possible small densities, may cause numerical (underflow) problems. Replacing $w^{1-\alpha}$ by the probability density function ${w^{1-\alpha}}/{\int w^{1-\alpha}}$, as suggested in \citet{picheny10}, can solve such difficulties. In practice, a  Monte Carlo method must be used to estimate the normalizing constant. 

%


For a DoE of dimension one or two, a Cartesian grid over the design space $\Omega$ can be used to solve the numerical integration and optimization problems \citep{picheny10}. In more general cases of higher dimension, stochastic integration and global optimization techniques should be preferred, e.g. Monte Carlo methods and SA algorithms (Appendix C). \\

\section{Numerical experiments}\label{numerical.experiments2}
\label{tests}

\textcolor{black}{In this section, numerical studies are conducted on a manageable example to assess the performances of both adaptive kriging strategies. The performances of the $\WIMSE$ and $\ECD$ criteria are compared with the standard {\it maximin}-LHD and the simple MMSE (maximum MSE) criterion, defined by
\begin{eqnarray*}
z_{(N+1)} & = & \argmin {z^* \in \Omega} \, \max_{z\in\Omega} \MSE\left(z|D_N \cup\{z^*\}\right),
\end{eqnarray*}
under the same evaluation budget. A good kriging metamodel has been built using a large DoE for playing a benchmark role.} \\



Consider the parametric function previously used in \citet{bastos09}:
\begin{eqnarray} \label{eq:modelToyTwo}
H(x_1,x_2) & = & \left(1-\exp\left(-\frac{1}{2x_2}\right) \right)\left(\frac{2300x^3_1+1900x^2_1+2092x_1+60}{100x^3_1+500x^2_1+4x_1+20} \right),
\end{eqnarray}
with $x_i\in[0,1], i=1,2$. In the experimental trials, the design domain $\Omega=[0,1]^2$. 
 The dataset $\mathbf{Y}=(Y_i,i=1,\dots,30)$ of size $n=30$ is simulated from the uncertainty model \eqref{modelToyTwo} where the missing data $X_i$ is generated with the following Gaussian distribution, truncated in domain $\Omega$:
\begin{eqnarray} 
X_i & \sim & \mathcal{N}_2\left\{\left(\begin{array}{c}  0.52 \\
                                   0.59 \\ 
                                  \end{array} \right),\left( \begin{array}{ll}  0.19^2 & 0 \\
                                   0 & 0.25^2 \\ 
                                  \end{array} \right)\right\} \cdot \1_{\Omega},
\end{eqnarray}
and the error term $U_i$ is the realization of a $\mathcal{N}_1(0, 10^{-5})$ random variable. Moreover, in \eqref{InfoPrior1} and \eqref{InfoPrior2}, the hyperparameters are chosen as follows:  
$a=1$, $\nu=5$, $\mu=(0,0)$ and 
\begin{eqnarray*}
\Lambda & = & 2\cdot\left( \begin{array}{ll}  0.18^2 & 0 \\
                                   0 & 0.4^2 \\ 
                                  \end{array} \right).
\end{eqnarray*}
%

\textcolor{black}{In practice, the burn-in period of the MCMC algorithm can be verified by the Brooks-Gelman diagnostic $\widehat{R}_{BG}$ of convergence \citep{brooks98}. It was calculated every 50 iterations and the convergence was not accepted until $\widehat{R}_{BG}<1.05$ for at least 3,000 successive iterations.}


The main features of the generated DoEs are summarized on Table \ref{table:DoE1}. All initial DoEs consist of the same five points produced by 
{\it maximin}-LHD, and then are completed by five other points selected by the criteria. \textcolor{black}{Table \ref{table:ToyModel} displays the value of parameters involved in carrying out the two criteria and the SA algorithm.}

\begin{figure}[!h]
\begin{minipage}[b]{0.5\linewidth}
      \centering \includegraphics[width=2in,height=1.8in]{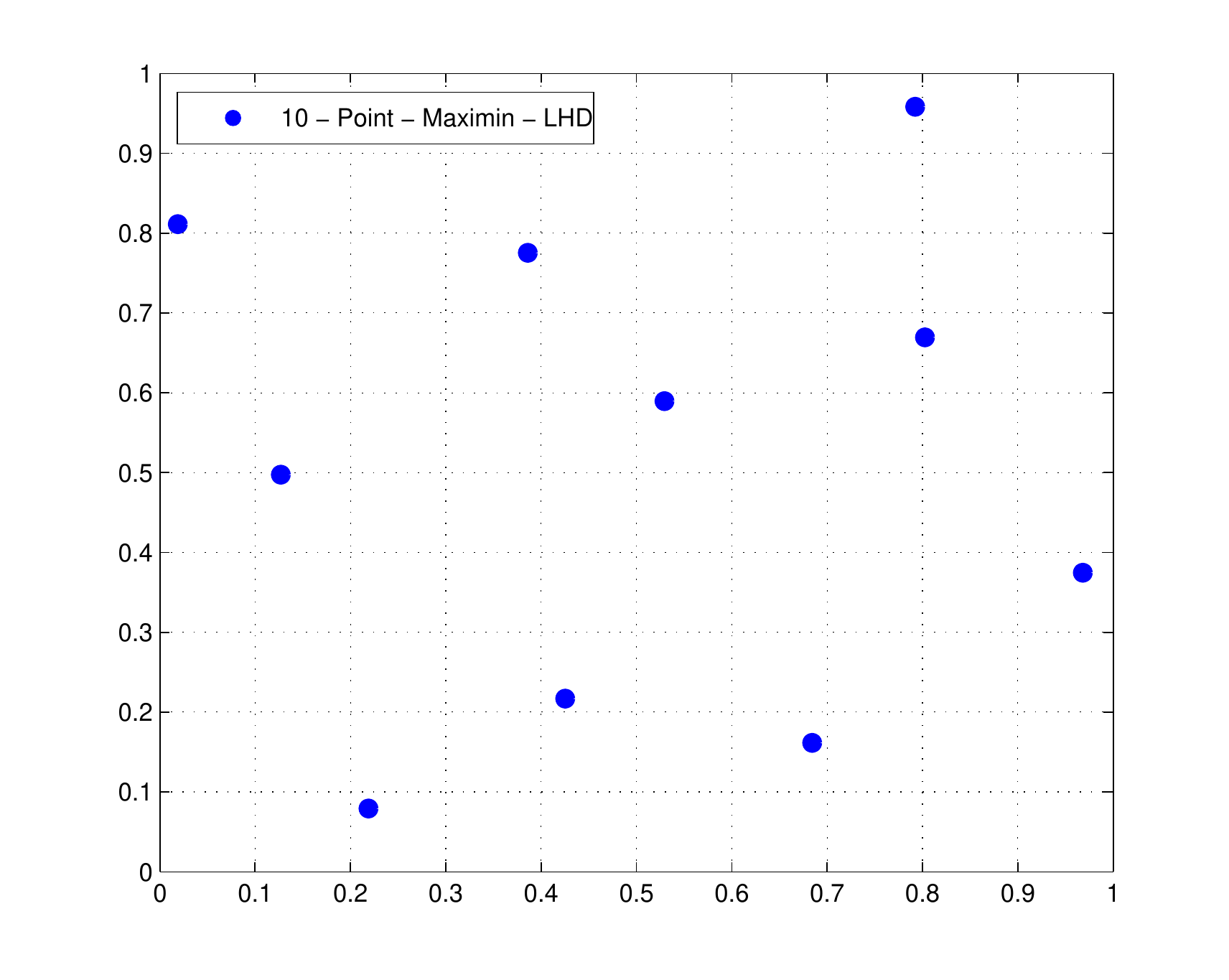}
   \end{minipage}\hfill
\begin{minipage}[b]{0.5\linewidth}
      \centering \includegraphics[width=2in,height=1.8in]{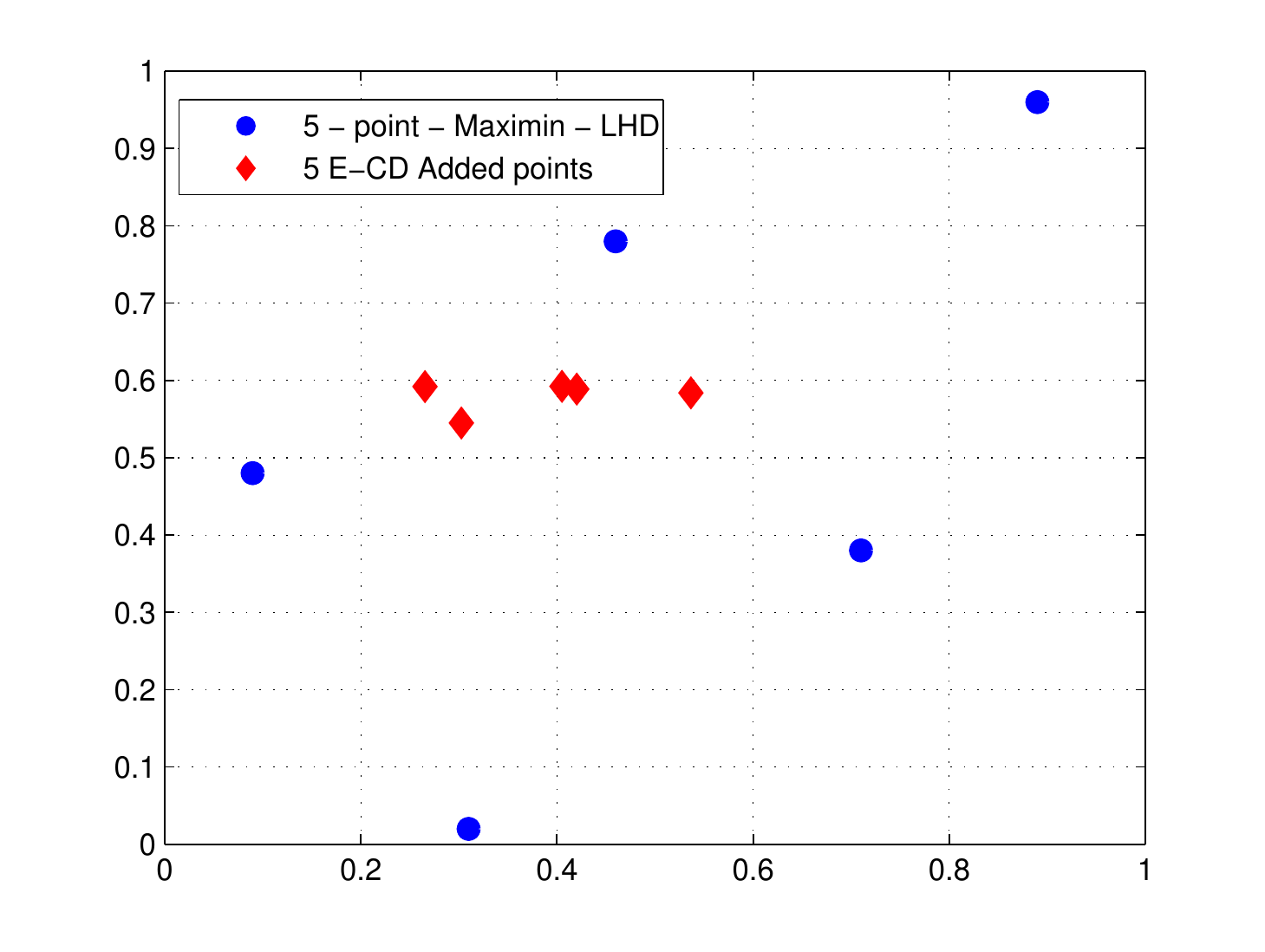}
   \end{minipage}
\\  
   \begin{minipage}[b]{0.3\linewidth}
      \centering \includegraphics[width=2in,height=1.8in]{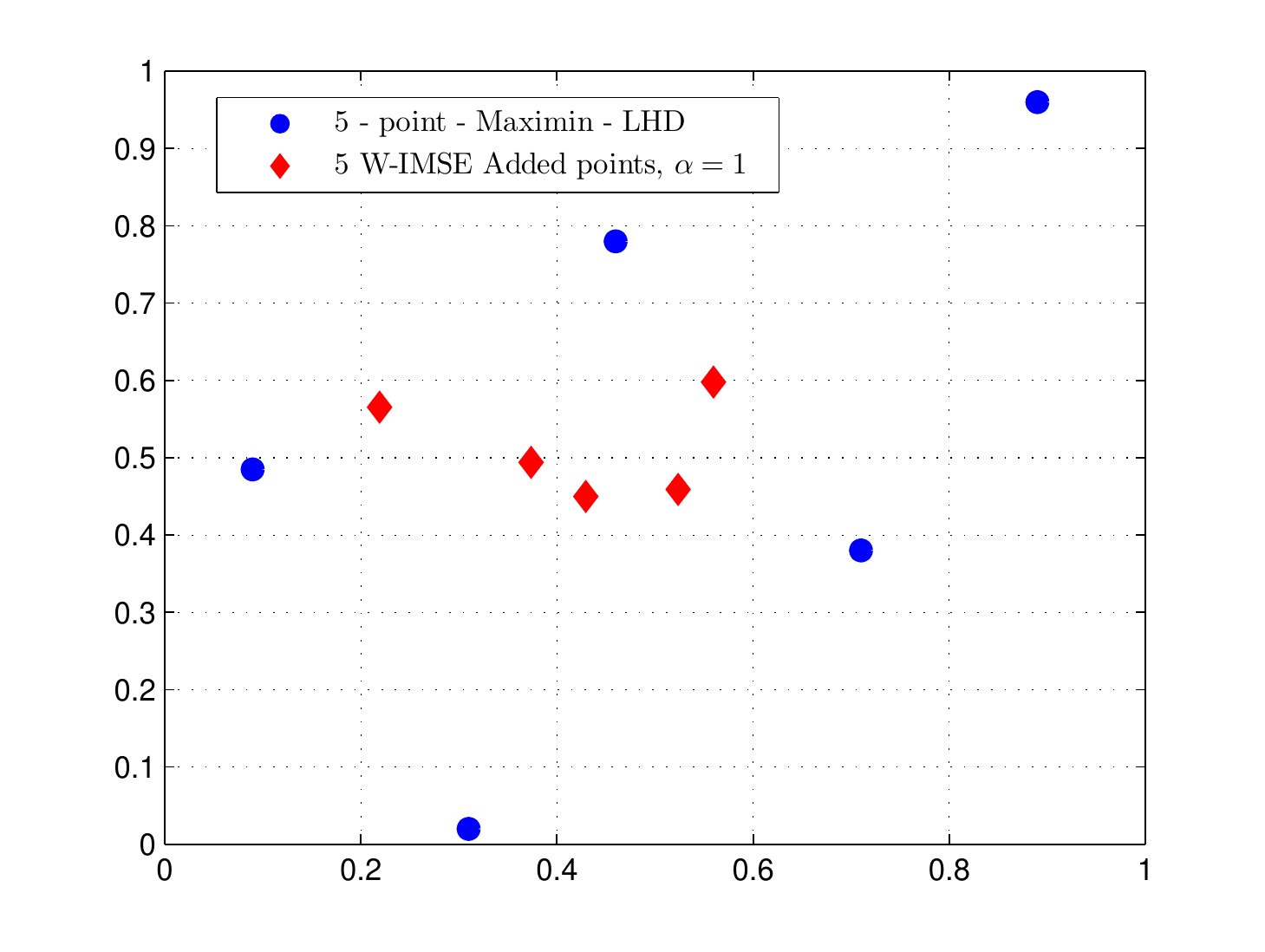}
   \end{minipage}\hfill   
   \begin{minipage}[b]{0.33\linewidth}
      \centering \includegraphics[width=2in,height=1.8in]{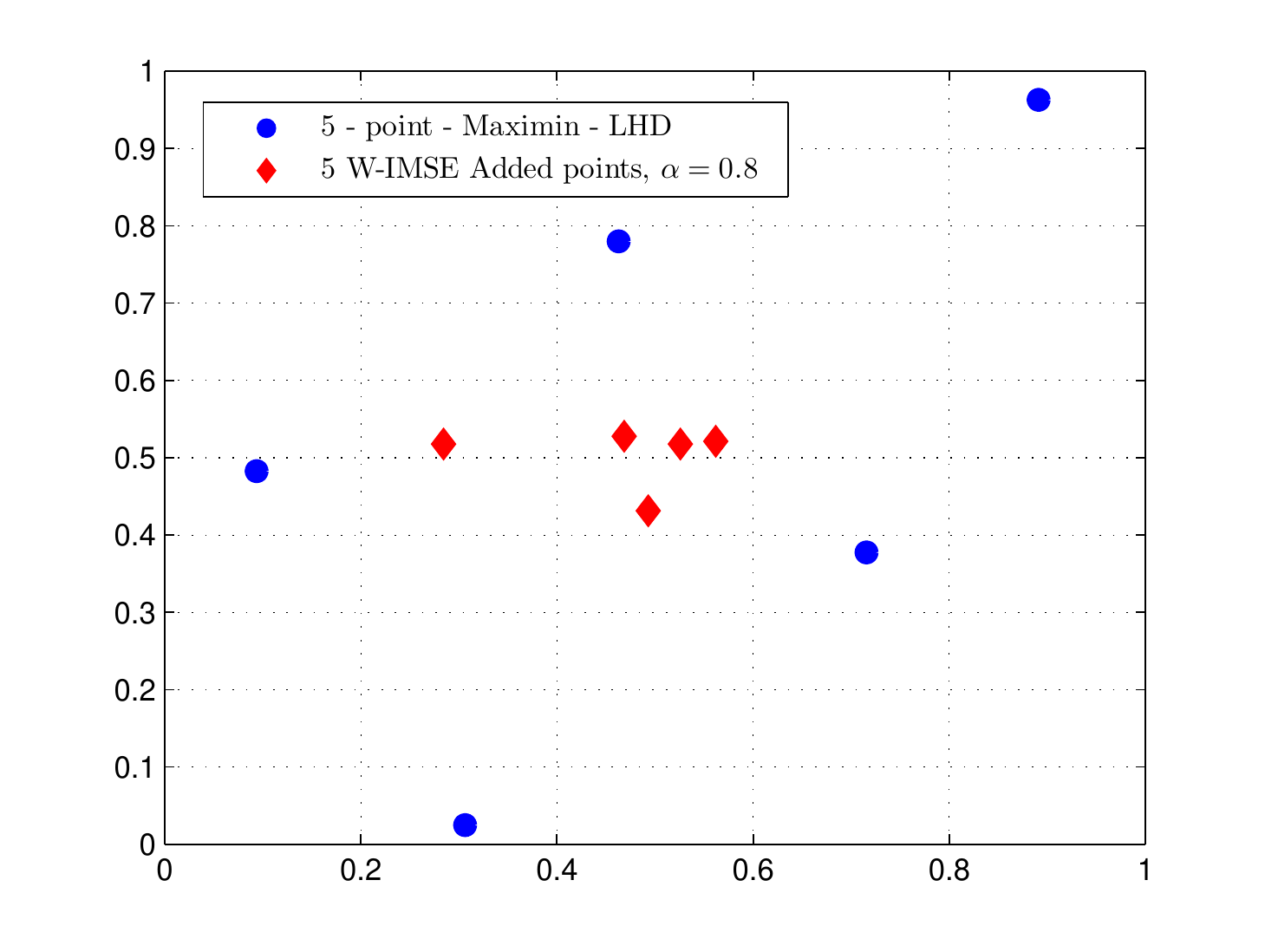}
   \end{minipage}
      \begin{minipage}[b]{0.3\linewidth}
      \centering \includegraphics[width=2in,height=1.8in]{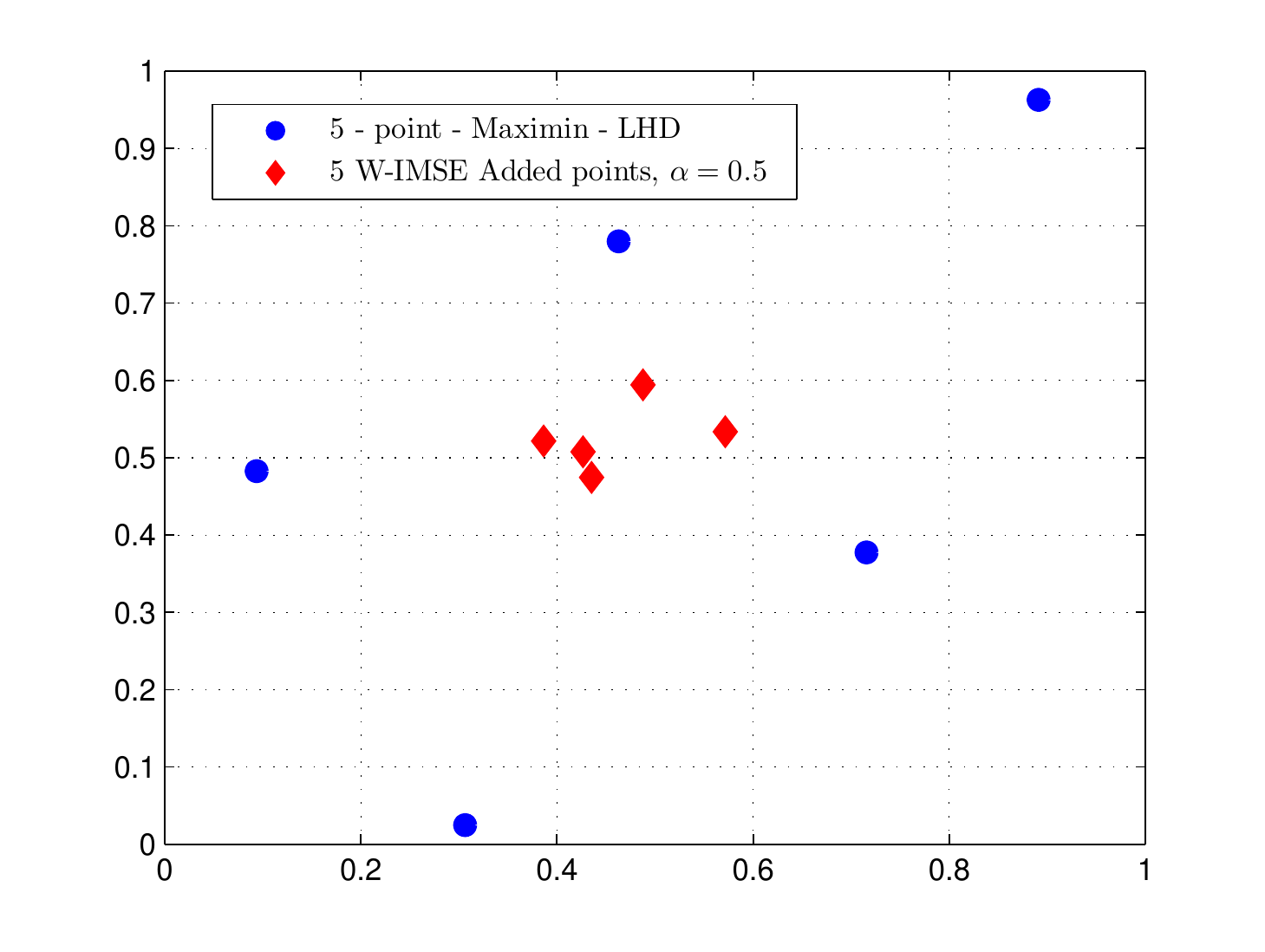}
   \end{minipage}
   \\
   \begin{minipage}[b]{0.5\linewidth}
      \centering \includegraphics[width=2in,height=1.8in]{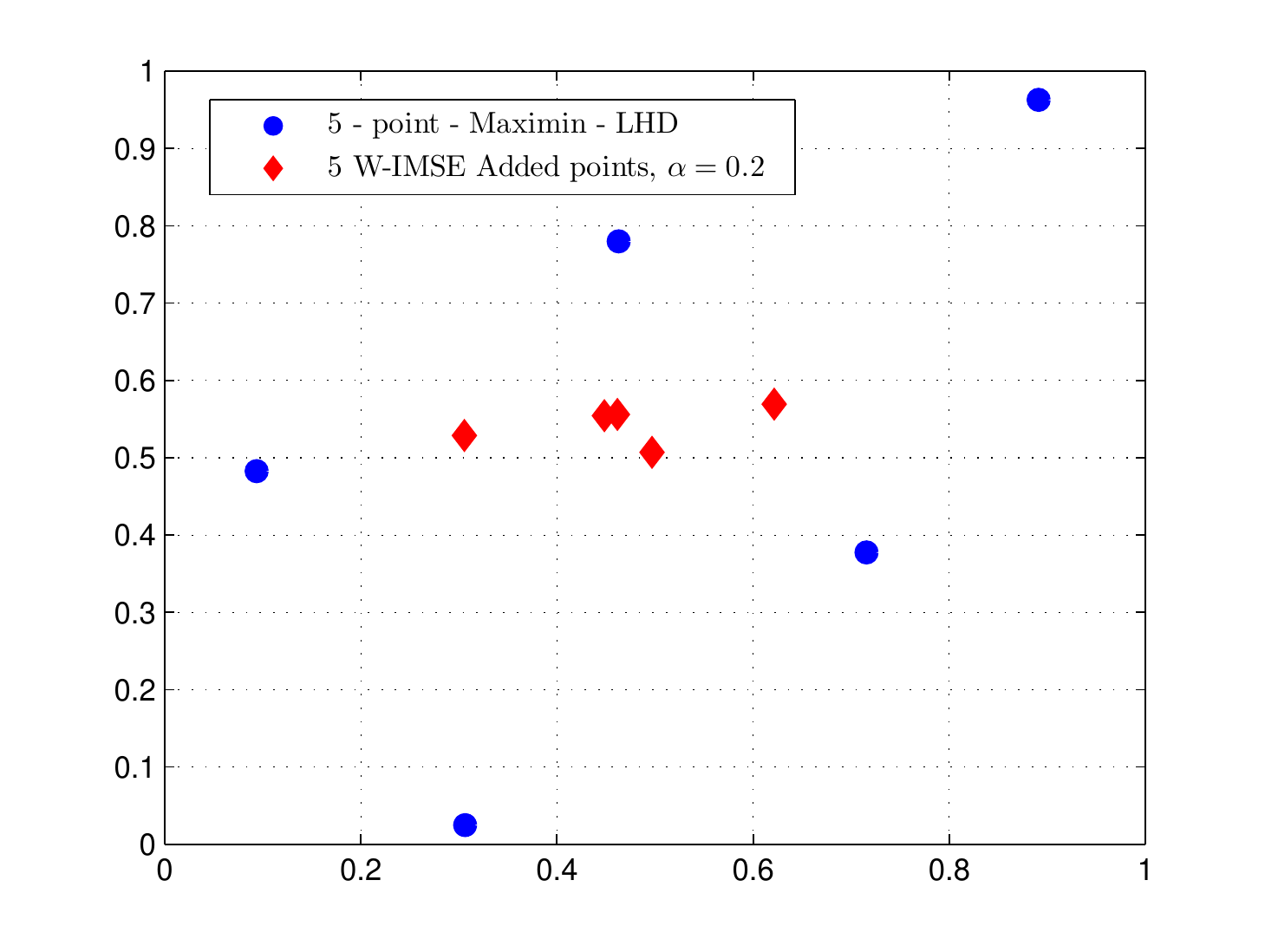}
   \end{minipage}\hfill
\begin{minipage}[b]{0.5\linewidth}
      \centering \includegraphics[width=2in,height=1.8in]{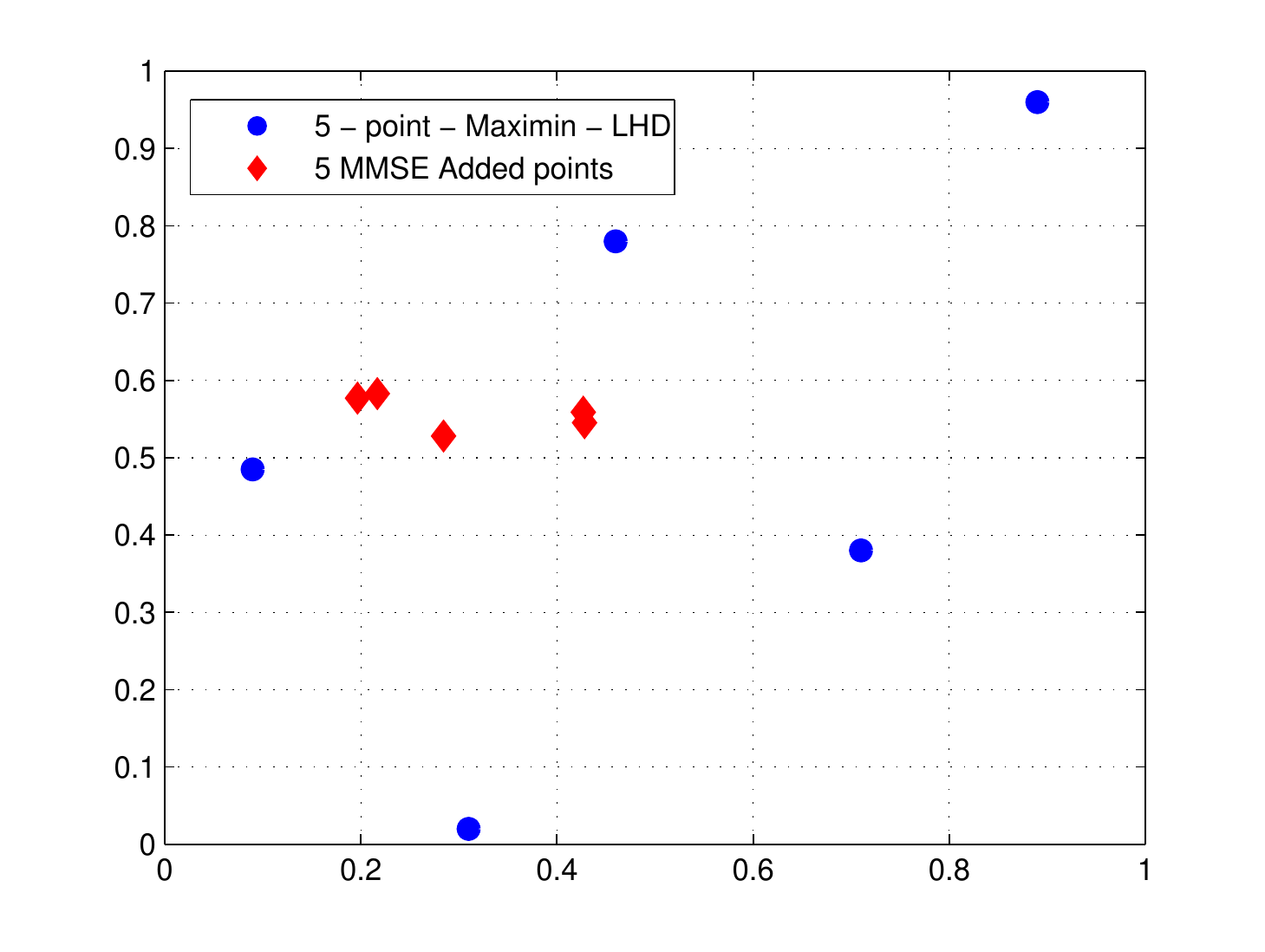}
   \end{minipage}

   \caption{Standard {\it maximin}-LHD, $\ECD$ design, $\WIMSE$ designs of experiments with $\alpha=1, 0.8, 0.5, 0.2$ and MMSE design (two-dimensional toy example).}\label{fig:DesignWIMSE2D}
 \end{figure}
\begin{figure}[!h]
      \includegraphics[scale=0.35]{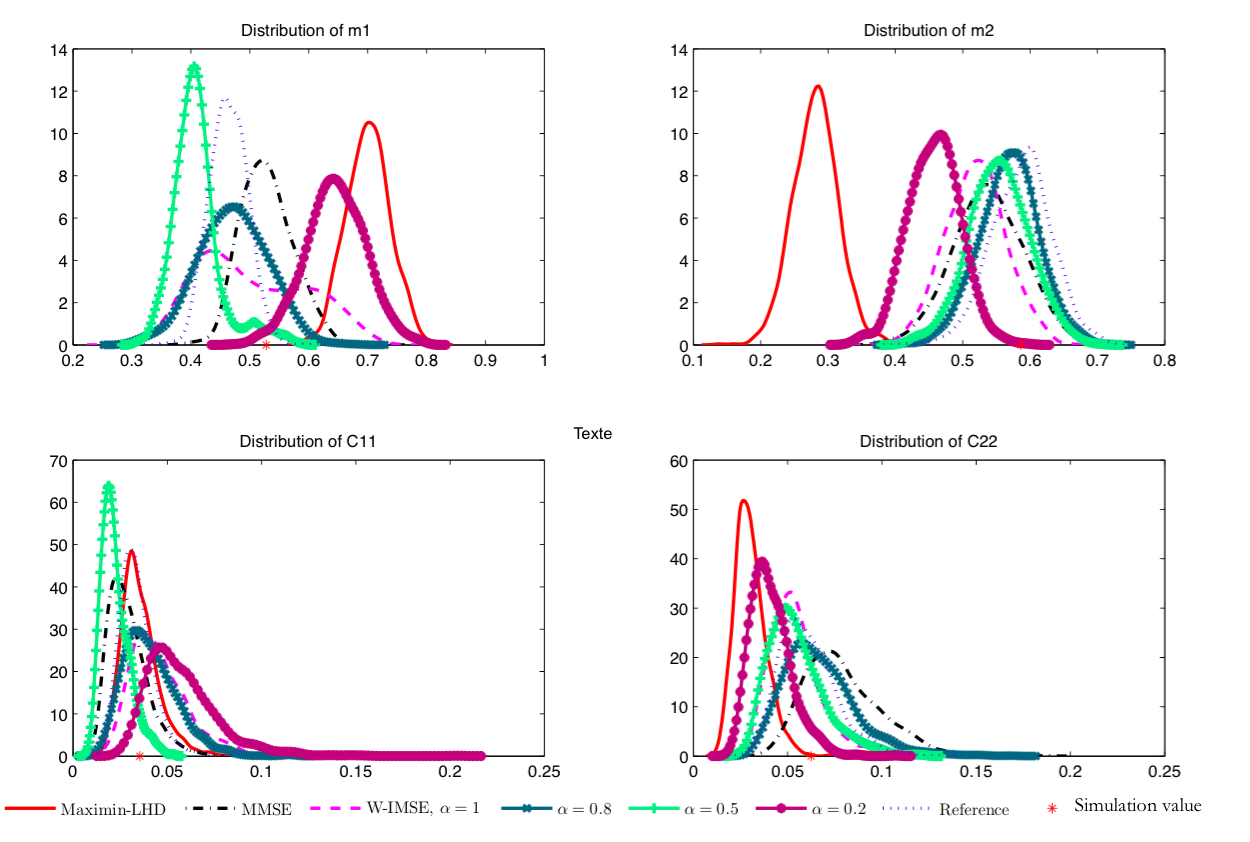}
        \caption{Posterior distributions of $\theta$ with benchmark, standard {\it maximin}-LHD, MMSE design and $\WIMSE$ designs with $\alpha=1, 0.8, 0.5, 0.2$  (two-dimensional toy example).}\label{fig:TestKrigWIMSE2D}
\end{figure}
\begin{figure}[!h]

       \includegraphics[scale=0.4]{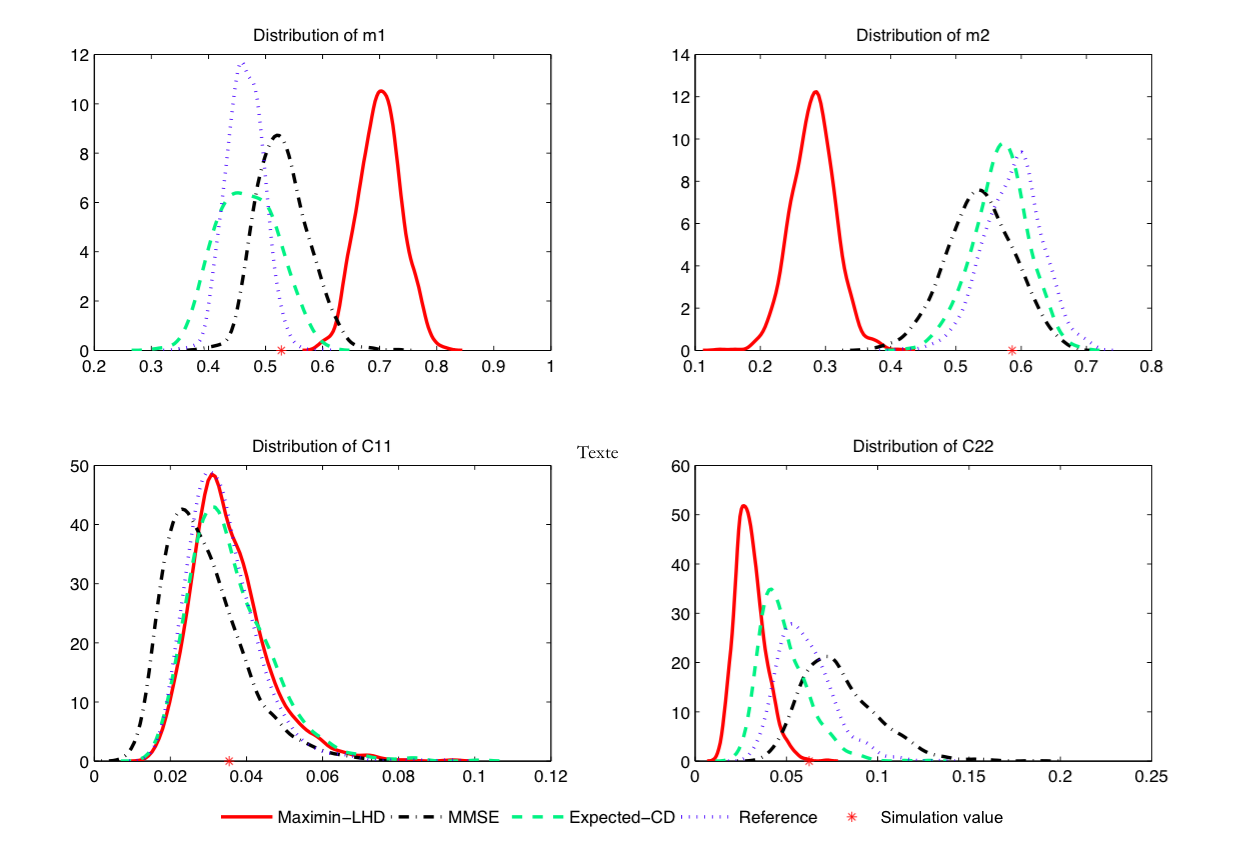}
        \caption{Posterior distributions of $\theta$ with benchmark, standard {\it maximin}-LHD, MMSE design and $\ECD$ design (two-dimensional toy example). }\label{fig:krigECDTwoTyo}
\end{figure}

Figure \ref{fig:DesignWIMSE2D} provides a comparison of all designs with the standard 10-points-{\it maximin}-LHD (encompassing the initial DoE). For the W-IMSE criterion, the added points are found not far from the hypothesized mean $(0.5,0.7)$ and the four $\WIMSE$ designs are quite similar. However, the posterior distributions of $\theta$ are quite sensitive to the choice of $\alpha$. Figure \ref{fig:TestKrigWIMSE2D} displays these posterior distributions for the corresponding metamodels. The $\WIMSE$ criterion improved the posterior distributions of $m_2$ and $C_{22}$, but the choices $\alpha=1, 0.5$ and $0.2$ do not work well for the posterior distribution of $m_1$ and $C_{11}$. It can be seen that the 10-points-{\it maximin}-LHD performs poorly, with respect to a 5-points-{\it maximin}-LHD sequentially completed. Moreover, the MMSE criterion performs correctly. However, other experiments, conducted using the best value $\alpha=0.8$ for the $\WIMSE$ criterion, are summarized on Figure \ref{fig:krigECDTwoTyo}. These results highlight, on this example, that the design build using the $\ECD$ criterion can significantly outperform the 10-point-{\it maximin}-LHD, can perform more efficiently than the MMSE criterion and can do as well as the  $\WIMSE$ criterion. 
\section{Case-study:  calibrating roughness coefficients of an hydraulic engineering model}\label{numerical.experiments}
 
\begin{figure}[!h]
      \centering \includegraphics[width=5in,height=2.5in]{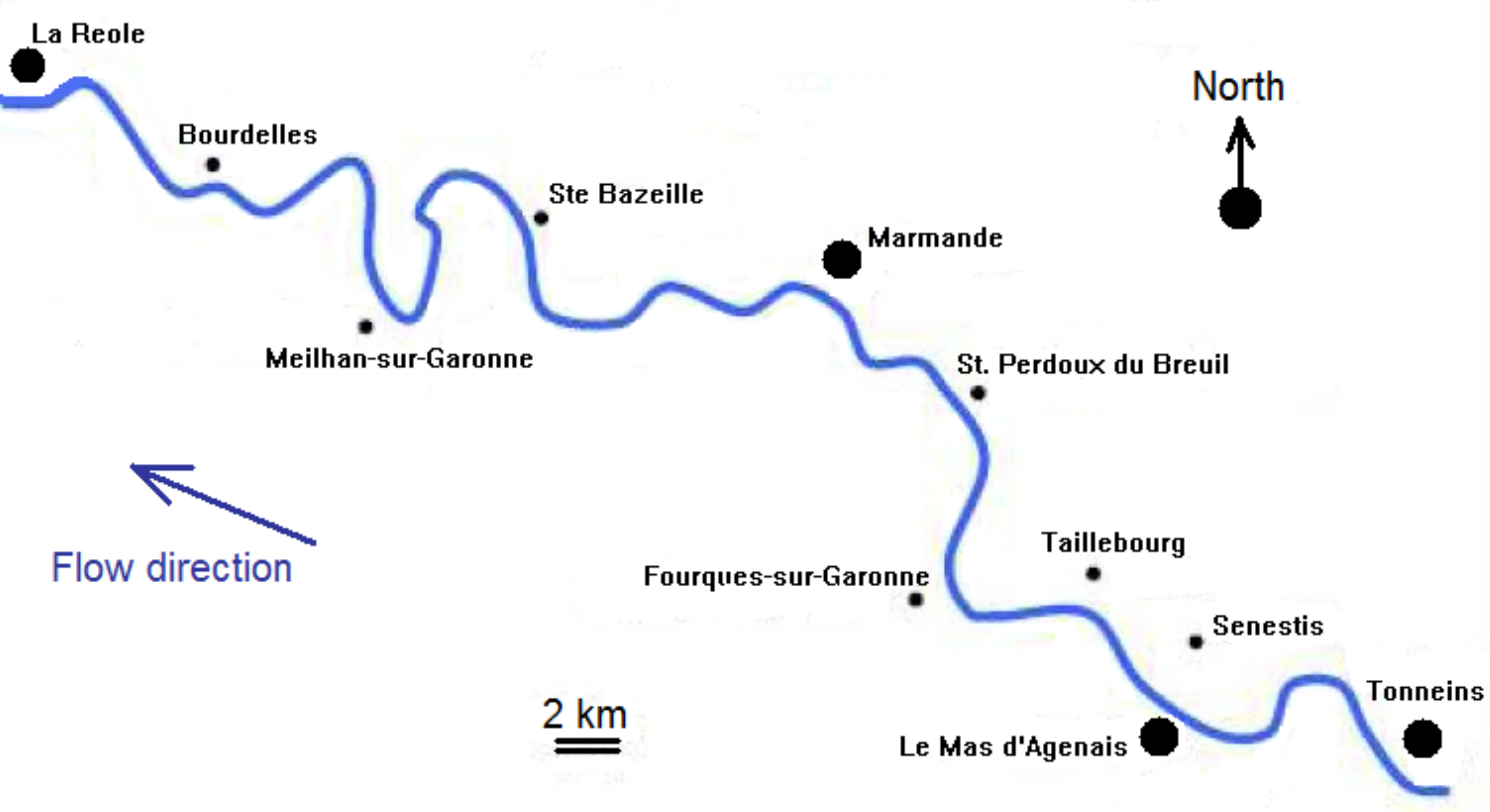}
      \caption{Riverbed profile of French river La Garonne.\label{fig:RiverGaronne}}
\end{figure}

\textcolor{black}{
The case-study that motivated this work is the calibration, from observed water levels $Y$ and upstream flow values $d$, of the roughness (so-called Strickler) coefficient $X$ of the hydraulic computer model TELEMAC-2D. This software tool is considered as one of the major standards in the field of 
free-surface flow by solving shallow water (Saint-Venant) equations \citep{galland1991}. This parameter vector summarizes the influence of the land nature on the water level, for a given discharge $d$.  The model is used here to reproduce in two dimensions (geographical coordinates) the downstream water level of the French river La Garonne between  {Tonneins and La R\'eole} (Figure \ref{fig:RiverGaronne}).} 

\begin{figure}[!h]
      \centering \includegraphics[width=5in,height=2.5in]{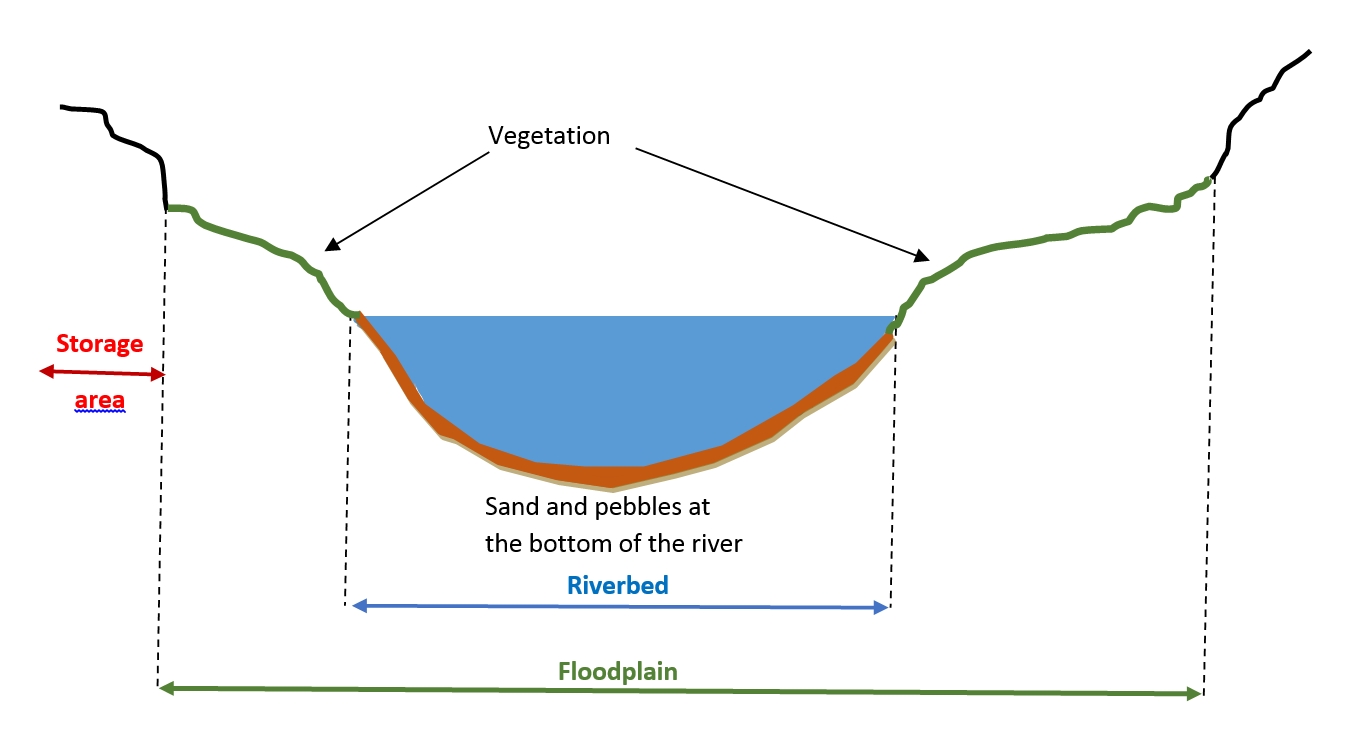}
      \caption{Cross-section of a classical river.\label{beds}}
\end{figure}

\textcolor{black}{
The flow simulation of this 50km  river section, including riverbed and floodplain (cf. Figure \ref{beds}), is conducted on very fine meshes defined by 41,000 knots, each parametrized by a roughness value. The dimension of $X$ is diminished to $q=4$ by taking account of: (a) the homogeneity of the land regularity in large areas surrounding the riverbed between four measuring stations (Table \ref{details} and Figure \ref{fig:RiverGaronne}) ; and : (b) the lack of observations of floodplain water levels at the uppermost subsection, which requires to fix the corresponding roughness coefficient. Details about the notation and meaning of each component of $X$ are provided in Table \ref{details}.}   \\

\begin{figure}[hbtp]
\centering
\includegraphics[width=12cm,height=8cm]{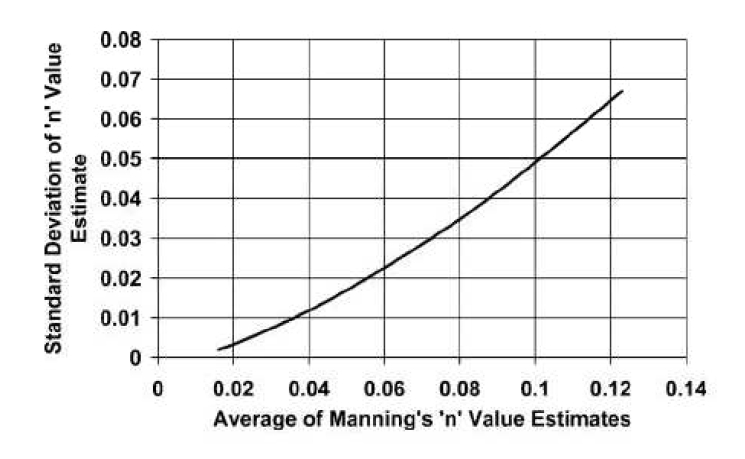}
\caption{Uncertainty over the estimators of Manning coefficient ($M=1/X$),  from \citet{USA96}. Cited (Fig. 3.5) in \citet{LIU09}.}
\label{usa9600}
\end{figure}

\textcolor{black}{
The strong but physically limited uncertainty that penalizes the knowledge of Strickler coefficients is compatible, according to \citet{WOH98}, with simple and classic statistical distributions as the Gaussian law (numerically truncated in 0).  Based on available bibliography summarized in Table \ref{prior.strickler} and after discussing with ground experts,  values for the hyperparameter $\mu$ for each dimension of $X$ were simple to elicit (see Table \ref{details}). It was more tricky to find information about the correlations between the $X$. The strong differences of land nature between the riverbed and the foodplain made plausible the assumption of independence between the corresponding components of $X$. On the contrary, it is likely that two connected riverbed section share roughness features. However, in absence of any additional information about these possibe correlations, $C_e$ was chosen diagonal:
\begin{eqnarray*}
C_{e}& = & \left( \begin{array}{llll}  \sigma^2_{\text{maj}} & 0 & 0 & 0 \\
                                   0 & \sigma^2_{\text{min}_{TA}} & 0 & 0 \\
                                   0 &  & \sigma^2_{\text{min}_{AA}} &  \\
                                   0 & 0 & 0 & \sigma^2_{\text{min}_{AL}} \\ 
                                  \end{array} \right).
\end{eqnarray*}
The calibration of each $\sigma$ was conducted by using marginal prior knowledge about the mean variation of the Manning coefficient $M=1/X$, discussed in \citet{LIU09} and displayed on Figure \ref{usa9600}. A  prior Manning estimator  $(\hat{M}=1/\mu,\sigma_M)$ can then be produced. A magnitude for the corresponding prior estimator of $\sigma$  (for the Strickler $X=1/M$) can be derived assuming that the results on Table \ref{prior.strickler} and Figure \ref{usa9600} summarize a large number of past estimations. Further to this assumption, a crude     in-law convergence 
\begin{eqnarray*}
\sigma^{-1}_M(\hat{M}-M) & \xrightarrow{{\cal{L}}}{} & {\cal{N}}(0,1).
\end{eqnarray*}
associated to a Delta method provides the approximate result
\begin{eqnarray*}
\mu^{-2}\sigma^{-1}_M(\mu-X) & \xrightarrow{{\cal{L}}}{} & {\cal{N}}(0,1),
\end{eqnarray*}
and finally $\sigma^2\simeq \mu^{4} \sigma^{2}_M$. The prior assessments of these variances are provided on Table \ref{prior.strickler}, assuming a virtual size $a=1$ for each dimension (see $\S$ \ref{prior.elicitation} for details). 
 }

\textcolor{black}{
The relevance of a metamodelling approach was acknowledged since each run of TELEMAC-2D can take several hours. {\it Maximin}-LHD designs were produced over the domain $\Omega$, defined for the input vector $z=(x,d)$ as   
\begin{eqnarray*}
\Omega & = & \Omega_{\text{maj}} \times \Omega_{\text{min}_{TA}} \times \Omega_{\text{min}_{AA}} \times\Omega_{\text{min}_{AL}} \times \Omega_d
\end{eqnarray*}
in function of the bounds of variation domains summarized in Table \ref{prior.strickler}: $\Omega_{\text{maj}}=[0,30]$ and $\Omega_{\text{min}_{TA}}=\Omega_{\text{min}_{AA}}=\Omega_{\text{min}_{AL}}=[20,70]$ (in $(m^{1/3}.s^{-1})$). The domain $\Omega_d$ was chosen as $[q_{0.05},q_{0.95}]=[510,2373]$ where $q_{\alpha}$ is the $\alpha-$order percentile of the known flow distribution, which is Gumbel with mode 1013 $m^3.s^{-1}$ and scale parameter 458.
} \\

\begin{figure}[!h]
      \centering \includegraphics[width=6in,height=4in]{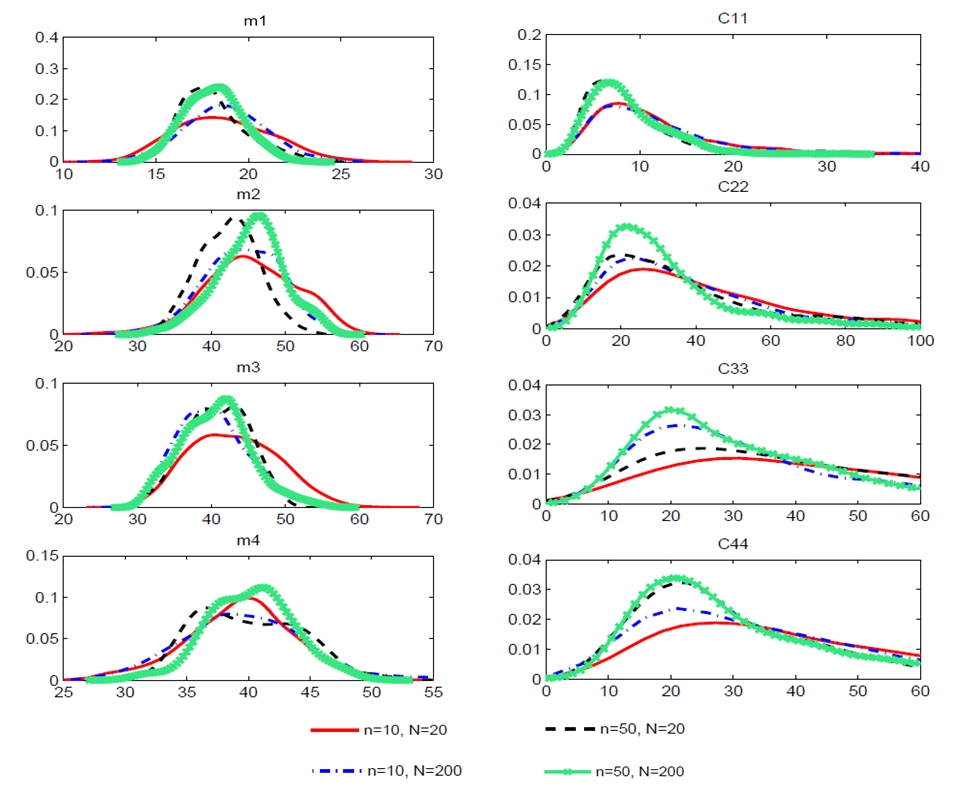}
      \caption{Approximations of the marginal posterior distributions of $\theta$ for several sizes $N$ {\it maximin}-LHD and two encompassed observation datasets ($n=10$ then $n=50$) using the MASCARET computer code. \label{mascaret-1}}
\end{figure}

\textcolor{black}{Before running TELEMAC-2D, however, a Bayesian inferential study was briefly conducted using the MASCARET simplified computer code  \citep{goutal12}, 
which describes a river by a curvilinear abscissa and uses the same input vector. While much more imprecise than TELEMAC-2D, the advantage of this simplified model is that the CPU time used for one run is shorter, so that the MCMC proposed in \cite{fu14} can be conducted in due time using a static {\it Maximin}-LHD design (and metamodelling calibrated once), using 20,000 iterations. The aim of this study was to test the agreement between the prior assessments and the observations, following recommendations in \cite{bousquet08,fu14}. A set of $n=50$ observations were available, among which the 10 most recent were preferentially selected, as the most representative of the actual conditions (riverbed homogeneity). For several sizes of design and the two datasets the marginal posterior distributions are displayed on Figure \ref{mascaret-1}. For each dimension, it appears that the regions of highest posterior density are in accordance with the prior guesses, which makes us confident in the relevance of the prior elicitation process. }\\

\textcolor{black}{Based on this good relevance of the Bayesian model, a comparison of the three designs considered in this article was conducted by comparing the {\it emulator errors} yielded by the designs, using the {\it coefficient of predictability} $Q_2$. A cross-validation {\it leave-one-out} version of this criterion is used here for computational simplicity \citep{vanderpoorten01}:}
\begin{eqnarray*}
Q_{2} & = & 1-\frac{\mbox{PRESS}}{\sum_{i=1}^{N} \big\|H(z_{(i)})-\overline H_{D_N}\big\|^2}.
\end{eqnarray*}
where $\overline H_{D_N}  = \frac{1}{N} \sum_{i=1}^N H(z_{(i)})$ and $\mbox{PRESS}  =  \sum_{i=1}^N e_{(i)}^2 \, = \, \sum_{i=1}^N \big\|H(z_{(i)})-\widehat{H}_{-i}(z_{(i)})\big\|^2$, with
\begin{itemize}
\item $e_{(i)}$ is the prediction error at $z_{(i)}$ of a fitted model without the point $z_{(i)}$;
\item $\widehat{H}_{-i}(z_{(i)})$ is the approximation of $H$ at $z_{(i)}$ derived from all the points of the design except $z_{(i)}$.
\end{itemize}

\textcolor{black}{The closer $Q_2$ to 1, the smaller the variance explained by the emulator and the better the quality of the design (in terms of prediction power for the metamodel). Four designs are tested. Two {\it Maximin}-LHD designs $D_{20}$ and $D_{500}$ of 20 and 500 points, respectively (the second one playing the role of a "reference design" leading to a very good approximation of the posterior distribution. Two other designs are sequentially elaborated using the ECD and WISE criterion, starting from an initial design $D_{10}$ of 10 points: 10 other points are added.} 

\begin{figure}[!h]
      \centering \includegraphics[width=4in,height=4in]{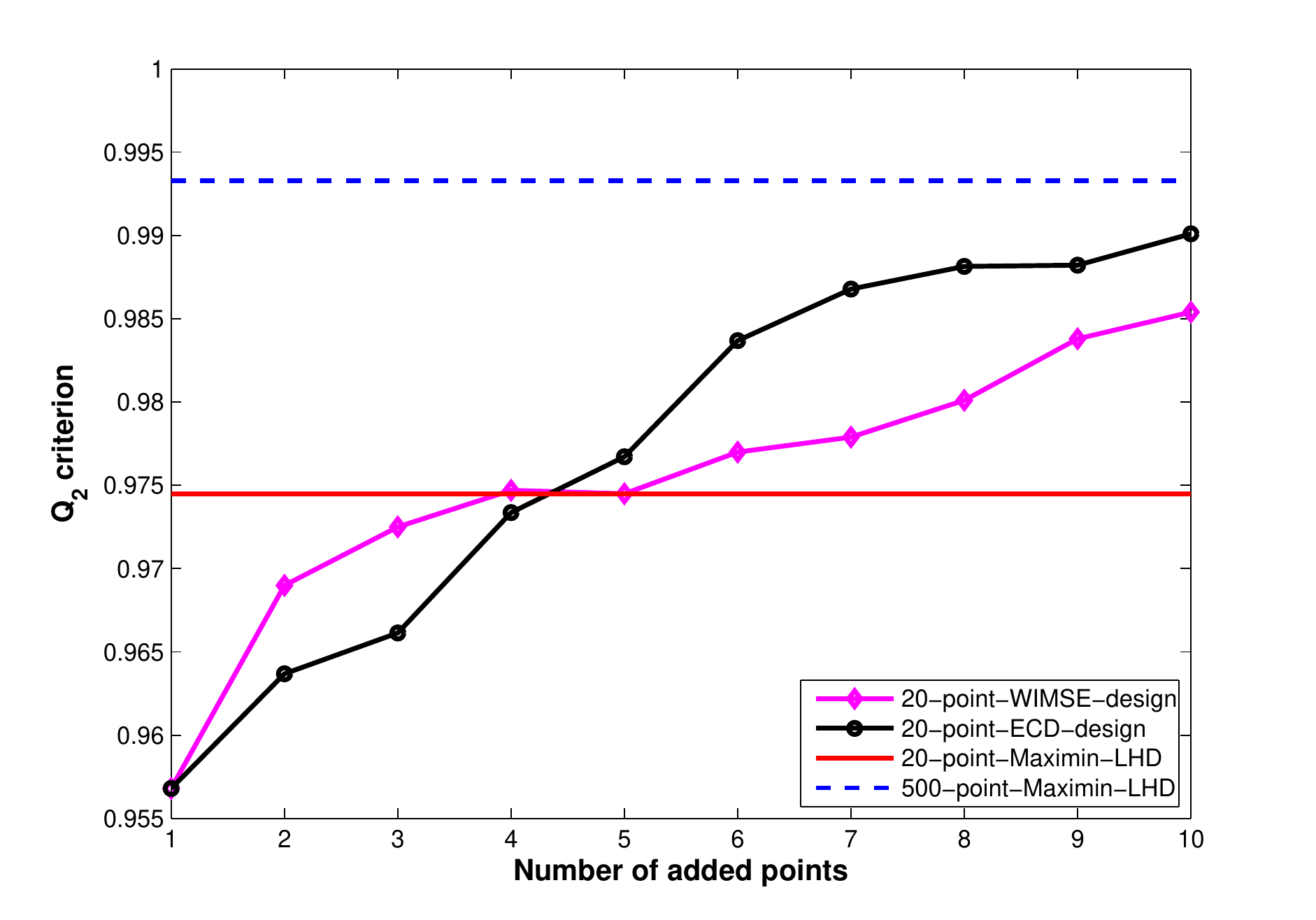}
   \caption{Comparison of the quality of different designs of numerical experiments using the MASCARET computer code.}\label{Q2compare}
 \end{figure}

\textcolor{black}{Displayed on Figure \ref{Q2compare}, the $Q_2$ coefficient related to the {\it maximin}-LHD $D_{20}$ equals 0.9745 and the benchmark $Q_2$ corresponding to the $D_{500}$ equals 0.9933. Starting from a design of 10 points only, it appears natural that other designs are characterized by a lower $Q_2$. However, by adding 10 points iteratively to the initial design $D_{10}$ according to the two proposed criteria, an increasing value of $Q_2$ is obtained, which quickly beats the predictability generated by the {\it maximin}-LHD $D_{20}$. Finally using the $\ECD$ criterion provides a slightly better $Q_2$ value than using the WISE criterion. } \\

\begin{figure}[!h]
      \centering \includegraphics[width=5in,height=5in]{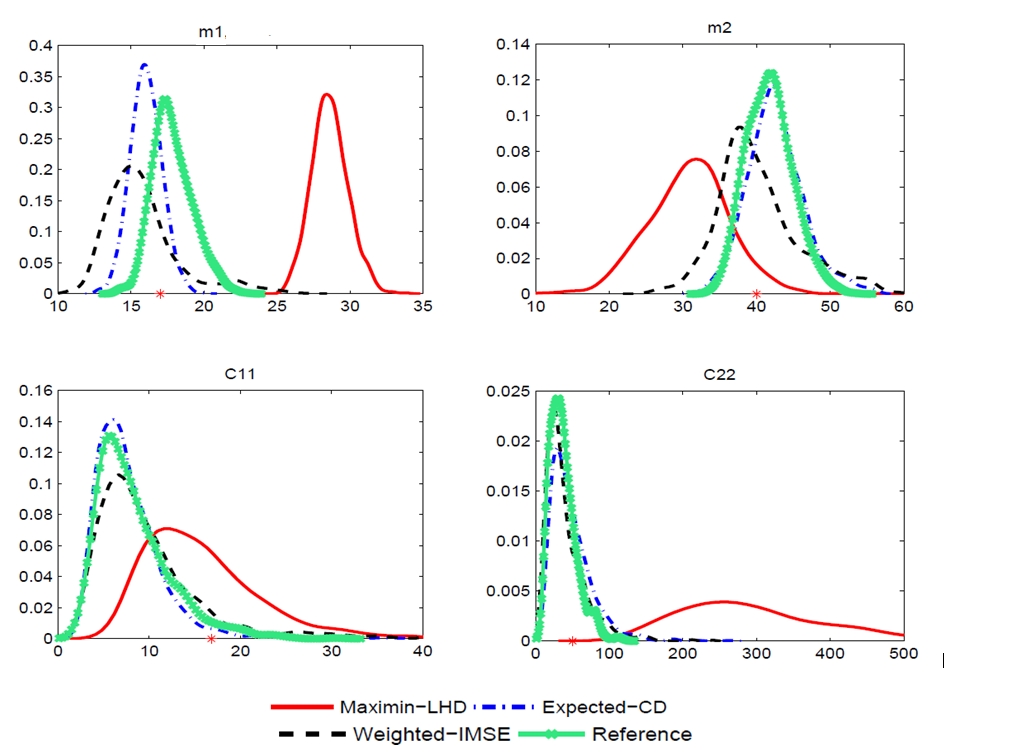}
   \caption{Approximations of the marginal posterior distributions of $\theta$ (first four dimensions) produced by several designs using the TELEMAC-2D computer code. The red stars indicates the prior means for each parameter.}\label{telemac-1}
 \end{figure}

\textcolor{black}{Coming back to the TELEMAC-2D computer code,  the convergence of MCMC chains were obtained (using the $n=10$ best observations) after 30,000 iterations. For the various designs proposed in this article, the marginal posterior distributions of the four first parameters are displayed on Figure \ref{telemac-1}. The  {\it Maximin}-LHD design $D_{20}$ (producing the approximate posterior in red) was made of 40 points, while other situations start from a DOE of 20 initial points, to which 20 other points are added sequentially (producing the approximate posteriors in blue and black).  The reference {\it Maximin}-LHD  design $D_{500}$ (producing the best approximation of the target posterior, in green) is made of 500 points, as for the MASCARET application. A better proximity of the approximate posterior distribution produced using $\ECD$ to the target can be again noticed with respect to the approximation produced by the WIMSE approach.}


\section{Conclusions and perspectives}
\label{conclusion}

This article aims to provide an adaptive methodology to calibrate, in a Bayesian framework, the distribution of unknown inputs of a nonlinear, time-consuming numerical model from observed outputs. This methodology is based on improving a space-filling design of experiments, typically the {\it maximin}-Latin Hypercube Design, that offers a non-intrusive exploration of the model. Kriging metamodelling is used to avoid costly runs of the model. 

In this methodology, two adaptive criteria have been proposed to  complete sequentially the current design. The first one is an adaptation of the standard Weighted-IMSE criterion to the Bayesian framework. It is obtained by weighting the $\MSE$ term over a region of interest indicated by the current full conditional posterior distribution. The other criterion, called Expected CD, is based on maximizing the Kullback-Leibler (KL) divergence between two consecutive approximate posterior distributions related to the DoE. A clearer interpretation can be given to the second criterion, as a crude approximation of the negative KL divergence between the target posterior and the current approximate posterior distributions.  

Numerical experiments have highlighted, on two examples, that applying this adaptive procedure can reduce the prediction error and improve the accuracy of the metamodeling approximation, compared with a standard space-filling DoE. Therefore such adaptive procedures appear to be useful when the CPU time required to compute an occurrence of the simulator $H$ of physical models is dramatically greater than the time required to run a Gibbs sampler, a Monte Carlo integration or to perform an optimization with a Simulated Annealing procedure. 

Both criteria involve expensive numerical integration. For a similar gain in information, the $\ECD$ criterion appears to be a little more expensive than the $\WIMSE$ criterion since it requires the calculation of the empirical $\KL$ divergence. However, in the definition of $\WIMSE$, the choice of $\alpha$ is quite important. As the second weight function is globally much smaller that the first prediction error, this balance parameter permits us to find a good behavior of this criterion. In this article, this important parameter was not systematically studied, but the computation of the best (or at least a "good") value of $\alpha$ makes the use of  $\WIMSE$ much less easy. In addition of this better interpretation (in information-theoretic terms), this feature lets us have a clear preference for the use of the ECD criterion. \\

This work is a first approach to designing sequential strategies for both exploring a black-box, time-consuming computer code  and in parallel calibrating some of its unobserved random inputs. The democratized use of metamodelling  requires, in practice, to make various approximations. For instance, it is current that the hyper-parameters of kriging metamodels are updated (e.g., by maximum likelihood estimation) after several additions of points to an original design, since each updating (which should be formally conducted after each addition of a new point) can be a costly operation itself without fundamental improvement \citep{toal08,toal11}. Following a same idea of reaching a trade-off between a theoretical aim and practical easiness, idealistic criteria are often necessarily approximated, or favored partially because their computation can made explicit. This is for instance the case of the Expected Improvement (EI) criterion proposed by \citet{jones98} which makes profit from the Gaussian properties of kriging metamodels. 

Such approximations appeared needed to conduct this first study and highlight the interest of the approach. The rationale developed in Section \ref{ECD} must now be followed by a truly theoretical work that could robustify the proposed choices, accompanied with more systematical simulation studies with other static or dynamic designs of numerical experiments. Especially, the statistical control of the metamodelling-based posterior approximation with respect to the target posterior should be a focus point in future studies, by making profit of the relationships between Kullback-Leibler divergences and discrepancy measures \citep{pollard13} as well as recent theoretical developments about relaxing assumptions under which metamodelling provides a fair approximation of the real numerical model (e.g., \citet{vazquez11}). Such works are currently being conducted. For the present time, it must be noticed that the approximate posterior distribution produced by the ECD approach can be considered as  a fast non-intrusive way of modelling an instrumental distribution, to be used in a final step of importance sampling (typically to compute a posterior mean), provided a small computational budget be kept or made available for running the numerical model.


\section*{Acknowledgments}

The authors gratefully thank Gilles Celeux (INRIA) for many fruitful discussions and advices. 
This work was partially supported by the French Ministry of Economy in the context of the CSDL (\textit{Complex Systems Design Lab}) project of the Business Cluster System@tic Paris-R\'{e}gion.

\bibliographystyle{apalike}
\bibliography{biblio}

\newpage

\subsection*{Appendix A. Metropolis-Hastings step within the Gibbs sampler\label{Appen:MH}}

At step $r+1$ of Gibbs sampling, after simulating $m^{(r+1)}$,$C^{(r+1)}$, the missing data $\mathbf{X}^{(r+1)}$ can be updated
with a Metropolis-Hasting ($\mbox{MH}$) algorithm. The MH step is updating $\mathbf{X}^{(r)}=(X^r_1, \dots, X^r_n)'$ 
in the following way:
\begin{itemize}
\item \texttt{For $i=1,\ldots,n$}
\begin{enumerate}
\item \texttt{Generate $\widetilde{X}_i \sim J(\cdot \mid X_i^r)$ where $J$ is the proposal distribution.}
\item \texttt{Let}
\begin{eqnarray*}
 \alpha(X_i^{r}, \widetilde{X}_i) & = & \min \Big(\frac{\pi_{\widehat{H}}(\widetilde{\mathbf{X}} \mid \boldsymbol{\mathcal{Y}}, \theta^{(r+1)},\rho,\mathbf{d},H_D)\, J(X_i^r|\widetilde{X}_i)}{\pi_{\widehat{H}}(\mathbf{X}^{(r)} \mid \boldsymbol{\mathcal{Y}}, \theta^{(r+1)},\rho,\mathbf{d},H_D)\, J(\widetilde{X}_i|X_i^r)}, 1\Big),\nonumber\\
\label{eq:alpha_2}
\end{eqnarray*}
where
\begin{eqnarray*}
\widetilde{\mathbf{X}} & = & \Big(X_1^{r+1},\, \dots,\,X_{i-1}^{r+1},\,\widetilde{X}_i,\,X_{i+1}^r,\,\dots,\,X_n^r \Big)'\\
\mathbf{X}^{(r)}  & = & \Big(X_1^r,\, \dots,\,X_{i-1}^r,\,X_i^r,\,X_{i+1}^r,\,\dots,\,X_n^r \Big)'
\end{eqnarray*}

\item \texttt{Take}
\begin{eqnarray*}
 X_i^{r+1} & = & \left\{\begin{array}{ll} \widetilde{X}_i & \textrm{\texttt{with probability} $\alpha(X_i^r, \widetilde{X}_i)$},\\
 X_i^{r+1} & \textrm{\texttt{otherwise.}}
\end{array}\right.
\end{eqnarray*}
\end{enumerate}
\end{itemize}
\paragraph{Remarks} 
\begin{itemize}
\item Many choices are possible for the proposal distribution $J$. 
It appears that choosing an independent MH sampler with $J$ chosen to be the normal distribution $\mathcal{N}\Big( m^{(r+1)}, C^{(r+1)}\Big)$
give satisfying results for the model (\ref{eq:model}).
\item In practice, it can be beneficial to choose the order of the updates by a random permutation of
$\{1, \ldots, n\}$ to accelerate the convergence of the Markov chain to its limit distribution. 
\end{itemize}

\subsection*{Appendix B. Nearest-Neighbor approach\label{Appen:NN}}

The  {Kullback-Leibler} (KL)
divergence between samples $\Theta^i$ and $\Psi$ can be empirically calculated through the Nearest-Neighbor approach. 
\begin{eqnarray}\label{eq:NNesti}
\widehat{\KL}_{L_1,L_2}(\Theta^i\, ||\,\Psi) & = & \frac{d}{L_1}\sum_{j=1}^{L_1} \log \frac{\nu_{L_2}(\theta^i_j)}{\rho^i_{L_1}(\theta^i_j)}\,+\,\log\frac{L_2}{L_1-1},
\end{eqnarray}
where $d$ denotes the dimension of the parameter $\theta$ ($2q$ in our case), $\nu_{L_2}(\theta^i_j)$ denotes the (Euclidean) distance between $\theta^i_j\in\Theta^i$ and its nearest neighbor in sample $\Psi$ 
\begin{eqnarray*}
\nu_{L_2}(\theta^i_j) & = & \min_{\substack{r=1,\dots, L_2}}\,||\theta_r-\theta^i_j||_2,
\end{eqnarray*}
and $\rho^i_{L_1}(\theta^i_j)$ denotes the (Euclidean) distance of  $\theta^i_j$ to its nearest neighbor in  $\Theta^i$ except itself (as it is also included in $\Theta^i$)
\begin{eqnarray*}
\rho^i_{L_1}(\theta^i_j) & = & \min_{\substack{l=1,\dots, L_1;\, l\neq j}}\,||\theta^i_l-\theta^i_j||_2.
\end{eqnarray*}

It has been proved in \cite{wang09} that under some regularity conditions on the
samples $\Theta^i$ and $\Psi$, the estimator $\widehat{\KL}_{L_1,L_2}(\Theta^i\, ||\,\Psi)$
is consistent in the sense that
\begin{eqnarray}
\lim_{L_1,L_2\rightarrow \infty} \mathbb{E}\left(\widehat{\KL}_{L_1,L_2}(\Theta^i\, ||\,\Psi) - \KL(\Theta^i\, ||\,\Psi) \right)^2 & = & 0,
\end{eqnarray}
and asymptotically unbiased, i.e.
\begin{eqnarray}
\lim_{\substack{L,R\rightarrow \infty}} \mathbb{E}\left[\widehat{\KL}_{L_1,L_2}(\Theta^i\, ||\,\Psi) \right] & = &  \KL(\Theta^i\, ||\,\Psi).
\end{eqnarray}

\subsection*{Appendix C. Simulated Annealing algorithm (searching for the minimum of a function $f$)
\label{Appen:SA}}

Proposed by \citet{kirkpatrick83}, the SA algorithm is a stochastic optimization algorithm.

\texttt{Given the current point $z^{(k)}$, at iteration $k+1$~:} \\

\begin{enumerate}
\item \texttt{Generate $\widetilde{z}\sim\mathcal{N}\Big(z^{(k)},\sigma^2\Big)$, with a certain fixed variance $\sigma^2$.}
\item \texttt{Let}
\begin{eqnarray*}
\lambda\Big(z^{(k)}, \widetilde{z}\Big) & = & \min\Big(1, \exp\Big(\frac{f(z^{(k)})-f(\widetilde{z})}{\beta_{k+1}}\Big) \Big),
\end{eqnarray*}
\texttt{where $\beta_{k+1}$ is the current temperature at step $k+1$. }
\item \texttt{Accept}
\begin{eqnarray*}
 z^{[k+1]} & = & \left\{\begin{array}{ll} \widetilde{z}, & \textrm{\texttt{with probability} $\lambda\Big(z^{(k)}, \widetilde{z}\Big)$},\\
 z^{(k)}, & \textrm{\texttt{otherwise.}}
\end{array}  \right.
\end{eqnarray*}
\item \texttt{Update $\beta_{k+1} = 0.99\times \beta_k$.}
\end{enumerate}

\clearpage


\begin{table}[h!] 
\small
\begin{center}
\begin{tabular}{lcc}
\hline
\textbf{DoE 1} & \multicolumn{2}{c}{10-point-{\it maximin}-LHD} \\
& \\
\textbf{DoE 2} & 5-points-{\it maximin}-LHD + & 5-points-$\WIMSE$ \\
& & or 5-points-$\ECD$\\
& & \hspace{0.1cm} or 5-points-MMSE\\
& \\
\textbf{DoE 3} & \multicolumn{2}{c}{100-points-{\it maximin}-LHD ({\it benchmark})} \\
\hline
\end{tabular}
\caption{Description of the three types of designs of experiments (DOE) (two-dimensional toy example).}
\label{table:DoE1}
\end{center}
\end{table}


\begin{table}[h!]
\small
\centering
\begin{tabular}{lccc}
\hline
\textbf{$\WIMSE$} & $\alpha$ & Number $L$ of iterations & Size $M$ of the        \\
                  &          & of the SA algorithm      & Monte Carlo algorithm  \\
                  & 1, 0.8, 0.5, 0.2&  1,000 & 1,000 \\
& \\
\textbf{$\ECD$}  & Number $M$ of  & Sizes $L_1$ and $L_2$ of            & Number $L$ of iterations \\
                 & generated GPs  & the samples $\Theta^i$ and $\Psi$   & of the SA algorithm \\
								 & 100 & 1,000 & 1,000 \\
& \\
\textbf{SA algorithm} & Initial point $x^{[0]}$ & Initial temperature $\beta$ & Standard deviation $\sigma$ \\
                      & $x$ & 100 & 100 \\
\hline
\end{tabular}											


\caption{Choice of parameters for the design criteria computation and the SA algorithm (two-dimensional toy example).}
\label{table:ToyModel}
\end{table}







\begin{table}[hbtp]
\centering
\begin{tabular}{ll}
\hline
Nature of surface & Value of Strickler coefficient ($m^{1/3}\cdot s^{-1}$) \\
\hline
& \\
{\bf Riverbed} & \\
Smooth concrete & 75-90 \\
Earthen channel & 50-60 \\
Plain river, without shrub vegetation & 35-40 \\ 
Plain river, with shrub vegetation & 30 \\
Slow winding natural river  & 30-50 \\
Very cluttered riverbed  & 10-30 \\
Proliferating algae  & 3.3-12.5 \\
& \\
{\bf Foodplain} & \\
Meadows, uncultivated fields & 20 \\
Cultivated lands with low size vegetation & 15-20 - {\bf 18} \\
Cultivated lands with large size vegetation & 10-15  - {\bf 13} \\
Bush and undergrowth areas  &  8-12 - {\bf 10} \\
Forest                  & $<$10 \\ 
Low density urban sprawl & 8-10 \\
High density urban sprawl & 5-8 \\ 
\hline 
\end{tabular}
\caption{Realistic ranges of value for the Strickler coefficient in function of the nature of the surface, summarized from \citet{USG89,WAL89,SEL97} and \citet{VIO98}. Median values in bold type are interpreted by international experts as the most likely values taking account of uncertainties about the nature of vegetation, topographic irregularities, etc. }
\label{prior.strickler}
\end{table}


\begin{table}[hbtp]
\centering
\begin{tabular}{c|llll}
 (Sub)section & Position & $X$ component & \multicolumn{2}{c}{Marginal hyperparameters $(m^{1/3}.s^{-1})$} \\
\hline
Tonneins     &  &  \\
  $\downarrow$ &  foodplain & $X_{s,\text{maj}}$ & $\mu_{\text{maj}}  =  17$ &  $\sigma_{\text{maj}} = 4.1$\\  
 La R\'eole    &             &                   &  \\ 	
& \\	
& \\
Tonneins     &  &  \\
$\downarrow$ &  riverbed & $X_{s,\text{min}_{TA}}$ & $\mu_{\text{min}_{TA}}   =  45$ &  $\sigma_{\text{min}_{TA}} = 7.1$ \\  
 Aval de Mas d'Augenais &             &            &  \\  
 $\downarrow$ &  riverbed  & $X_{s,\text{min}_{AA}}$  & $\mu_{\text{min}_{AA}}  =  38$ & $\sigma_{\text{min}_{AA}} = 7.1$ \\  
  Amont de Marmande &             &            &  \\                              
 $\downarrow$ & riverbed   &  $X_{s,\text{min}_{AL}}$  & $\mu_{\text{min}_{AL}}   =  40$ & $\sigma_{\text{min}_{AL}} = 7.1$  \\ 		
La R\'eole &             &            &  \\										
\hline
\end{tabular}
\caption{Detailed meanings and prior modelling for each component of $X$ (La Garonne roughness coefficients). The riverbed roughness coefficients are differentiated between the measuring stations listed in the first column. A virtual size $a=1$ was chosen for each dimension.}
\label{details}
\end{table}




\end{document}